\documentclass[a4paper,11pt]{article}

\usepackage{amsmath}
\usepackage{amsfonts}
\usepackage{amssymb}         
\usepackage{amsopn}
\usepackage{amsthm}
\usepackage{graphicx}
\usepackage{epsfig}
\usepackage{times}
\usepackage{verbatim}
\usepackage{mathtools}
\usepackage{subcaption}
\usepackage{authblk}
\usepackage{hyperref}
\usepackage[left=1.0in,top=1.2in,right=1.0in]{geometry}
\usepackage{enumerate}
\usepackage{algorithm}
\usepackage{algorithmic}
\usepackage{rotating}
\usepackage{hyperref}
\usepackage{qcircuit}
\usepackage{color}
\usepackage{lineno}

\newcommand{\bra}[1]{\left\langle{#1}\right|}
\newcommand{\ket}[1]{\left|{#1}\right\rangle}
\newcommand{\red}[1]{{{#1}}}

\newcommand*\samethanks[1][\value{footnote}]{\footnotemark[#1]}

\title{\red{Pareto}-Efficient Quantum Circuit Simulation Using Tensor Contraction Deferral\thanks{An earlier version of this paper was called ``Breaking the 49-Qubit Barrier in the Simulation of Quantum Circuits''}}

\author[1]{Edwin Pednault\thanks{Corresponding author; \url{pednault@us.ibm.com}}}
\author[1]{John A.~Gunnels\thanks{These authors contributed equally.}}
\author[1]{Giacomo Nannicini\samethanks}
\author[1]{Lior Horesh}
\author[2]{Thomas Magerlein}
\author[3]{Edgar Solomonik}
\author[4]{Erik W.~Draeger}
\author[4]{Eric T.~Holland}
\author[1]{Robert Wisnieff}
\affil[1]{IBM T.J.~Watson Research Center, Yorktown Heights, NY}
\affil[2]{Tufts University, Medford, MA}
\affil[3]{Dept.~of Computer Science, University of Illinois at Urbana-Champaign, Champaign, IL}
\affil[4]{Lawrence Livermore National Laboratory, Livermore, CA}

\date{}

\begin{document}


\maketitle


\begin{abstract}
  With the current rate of progress in quantum computing technologies,
  systems with more than 50 qubits will soon become reality. Computing
  ideal quantum state amplitudes for circuits of such and larger sizes
  is a fundamental step to assess both the correctness,
  performance, and scaling behavior of quantum algorithms
  and the fidelities of quantum devices.  However, resource
  requirements for such calculations on classical computers grow
  exponentially. We show that deferring tensor contractions can extend
  the boundaries of what can be computed on
  classical systems.  To demonstrate this technique,
  we present results obtained from a calculation of
  the complete set of output amplitudes of a universal random circuit
  with depth 27 in a 2D lattice of $7 \times 7$ qubits, and an
  arbitrarily selected slice of $2^{37}$ amplitudes of a universal
  random circuit with depth 23 in a 2D lattice of $8 \times 7$
  qubits. Combining our methodology with other decomposition
  approaches found in the literature, we show that we can simulate $7
  \times 7$-qubit random circuits to arbitrary depth by leveraging
  secondary storage. These calculations were thought to be impossible
  due to resource requirements. 
\end{abstract}

\section{Introduction}
In the last few years, significant technological progress has enabled
quantum computing to evolve from a visionary idea
\cite{feynman1982simulating} to reality \cite{NatureNewsIBMQ,
  castelvecchi2017quantum}. With existing devices now providing
$20$--$50$ qubits with controllable couplings, the potential of
quantum devices to perform computational tasks beyond what any
classical computer can perform is receiving much attention
\cite{harrow2017quantum,preskill2012quantum,boixo2018supremacy,aaronson2016complexity,farhi2016quantum}.
Among the many issues faced in advancing quantum computing technology,
two are particularly relevant for this paper: (1) the ability to
assess the correctness, performance, and scaling behavior of quantum
algorithms \cite{Gheorghiu2017verification}, and (2) the ability to
quantify circuit fidelity (e.g.,
\cite{bishop2017volume,reed2010high,gambetta2015building,boixo2018supremacy,linke2017experimental}). Fundamental
to both is the ability to calculate quantum state amplitudes for all
or arbitrarily selected portions of the quantum state---a task whose
difficulty grows exponentially in the size of the circuit, in
general. This paper presents a new methodology for this task, suitable
for calculating entire quantum states, that can handle circuits of
much larger size than those that had been presented in the
literature prior to an early preprint of this paper
  \cite{pednault2017breaking}.  Our approach to calculating quantum
amplitudes uses a tensor representation \cite{joshi1995matrices},
combining the flexibility of tensor computations with tensor slicing
methods and optimized secondary storage techniques.  A first stage of
simulation enables final quantum states of limited-depth circuits to
be calculated in slices instead of having to materialize entire
quantum states in memory.  A second stage of simulation enables such
limited-depth calculations to be extended to arbitrary depths by
leveraging secondary storage in cases where sufficient resources are
available.

A key building block of our approach is {\em tensor contraction
  deferral}. This technique allows non-adjacent contractions, in
addition to the adjacent contractions used in conventional
tensor-network-based circuit simulation algorithms (e.g.,
\cite{markov2008simulating}). Contraction deferral enables the
decomposition of circuits into subcircuits that can be simulated
independently and then combined to calculate the final quantum
amplitudes, reducing resource requirements. It is the central
technique of the first preprint of this paper
\cite{pednault2017breaking}. To demonstrate our methodology, we apply
it to {\em universal random circuits} constructed according to the
rules described in \cite{boixo2018supremacy}.  These circuits can be
embedded in a 2D lattice of qubits with nearest-neighbor couplings,
and are restricted to gates belonging to the set $\{H, CZ, X^{1/2},
Y^{1/2}, T\}$.  The depth of such a circuit is the number of layers
that the circuit can be partitioned into, in such a way that at any
given layer at most one gate acts on any given qubit.  As in
\cite{boixo2018supremacy}, the depth count does not include the first
layer of Hadamard gates, nor the final layer of Hadamard gates in the
revised benchmarks of \cite{markov2018quantum,villalonga2018flexible}.

We simulate two quantum circuits constructed according to the rules
described in \cite{boixo2018supremacy}: one with depth 27 for a $7
\times 7$ grid of qubits, the other with depth 23 for an $8 \times 7$
grid of qubits.  Simulations were performed in September 2017 at
Lawrence Livermore National Laboratory on the Vulcan supercomputer, an
IBM Blue Gene/Q system.  The primary data structures employed in the
depth-27 simulation require just over 4.5 Terabytes of main memory to
store results, and just over 3.0 Terabytes for the depth-23
simulation. In contrast, state-of-the-art techniques in existence at
the time \cite{haner2017simulation} would have required 8 Petabytes
and 1 Exabyte, respectively. Recent work
\cite{boixo2017simulation,chen201864,li2018quantum,chen2018classical,markov2018quantum,villalonga2018flexible,chen2019teleport,villalonga2019frontier,guo2019PEPS,zhang2019classical}
has improved upon our results and simulated larger circuits. A
majority of these papers use tensor contraction methods of some form,
which we analyze from the point of view of contraction deferral in
order to provide a unifying perspective. Although we do not simulate
circuits constructed according to the revised rules described in
\cite{markov2018quantum,villalonga2018flexible}, we use our approach
to provide a quantitative analysis of the impact of the rule changes
on the difficulty of simulation. We also present extensions of our
simulation technique in order to combine it with the decomposition
approach discussed in \cite{haner2017simulation}. The goal of these
extensions is the simulation of quantum circuits with greater depth
than discussed above. We show, for example, that $7 \times 7$ circuits
can in principle be fully simulated with all amplitudes calculated to
arbitrary depth in reasonable time (e.g., less than a day for a
depth-83 circuit) by leveraging secondary storage. We do not test
these extensions computationally, but we describe the foundations of
the methodology and estimate the computational resources that these
experiments would require. The simulation of $7 \times 7$ circuits to
arbitrary depth was previously thought to be out of reach
\cite{boixo2018supremacy,li2018quantum,chen2018classical}.




\section{Methods}
The majority of this paper is devoted to a high-level description of
the simulation methodology introduced in this paper; further details
are provided in the Supplementary Information.

\subsection{Simulation methodology: preliminaries}
The input of a simulation algorithm is a description of a quantum
circuit, and a specification of a set of amplitudes (i.e.,
coefficients of the quantum state obtained at the end of the circuit)
that have to be computed; we are mainly interested in the case in
which all amplitudes have to be computed, but we will address the case
of single-amplitude calculations as well. To test and debug the
implementation of a quantum algorithm, having access to the entire
quantum state via exact simulation is preferable to computing
amplitudes for specific (e.g., measured) outcomes, because this allows
the identification of properties of the outcome distribution that
cannot be inferred from a small number of samples.

Our analysis considers a hierarchy of storage devices, which has two
levels in its simplest form: primary storage, i.e., RAM, and secondary
storage, i.e., disk. Descending the hierarchy increases available
space but decreases access speed: the performance of a simulation
strategy should take into account space occupancy and number of
accesses at all levels of the hierarchy. This distinction is crucial
from a practical point of view. For example, the full quantum state of
a $49$-qubit system requires $2^{49}$ complex numbers (8 PB in double
precision); this eliminates the possibility of keeping the entire
state in primary storage, but does not rule this out for secondary
storage. Thus, disk usage must be allowed in order to have enough
space to store the simulation output for circuits of this (or larger)
size.  To do so, computations must be reorganized to minimize
  disk access.  The main numerical experiments discussed in this
paper employ only main memory to store intermediate results, but the
Supplementary Information describes how this approach can be extended so that
disk can be used to simulate $7 \times 7$-qubit and potentially
larger circuits to arbitrary depth. It is well known that a
single amplitude can be computed using linear space and
exponential time (i.e., number of floating point operations) using the
Feynman path approach, see e.g.,
\cite{aaronson2016complexity}. However, the time requirement of such
an approach grows to intractable levels very quickly
with increasing circuit size. We desire exact simulation algorithms
with manageable time requirements (i.e., hours -- not days), which
from a practical standpoint implies a small number of floating point
operations per amplitude per gate.  This requirement leads us to
  consider how to best exploit memory hierarchies.

In very broad terms, our approach consists of partitioning a quantum
circuit into subcircuits that can be simulated independently and then
recombined to obtain the desired result. Of course, a quantum circuit
cannot in general be split into subcircuits to be simulated
independently due to entanglement. However, when the action of the
circuit is represented in a purely algebraic manner, e.g., using a
graphical model known as a tensor network
\cite{joshi1995matrices,markov2008simulating}, it can be verified that
arbitrary splitting into subcircuits can be performed, although it may
require additional memory and/or computation to account for
entanglement between subcircuits. As a consequence, not all
decompositions of a circuit into subcircuits are memory-efficient
and/or computationally efficient.

\begin{figure}[tb]
  \centering
  \includegraphics[width=0.35\textwidth]{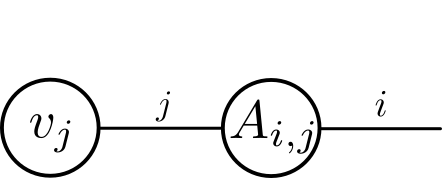}
  \caption{Example of a tensor network for the expression $\sum_j
    A_{i,j} v_j$.}
  \label{fig:tensor_network_example}
\end{figure}

A description of our simulation algorithm requires some preliminary
discussion regarding tensor networks. A {\em tensor} is a multilinear
map (i.e., a multidimensional generalization of a matrix), equipped
with a set of indices to address its elements; each index varies
within a specified range, which we always assume to be the set $\{0,
1\}$ in this paper --- corresponding to the basis elements $\ket{0},
\ket{1}$ for the vector space $\mathbb{C}^2$ in which the state of a
single qubit lives. The number of indices of a tensor is called its
{\em rank}. It follows that a rank-$k$ tensor with ranges $\{0, 1\}$
requires $O(2^k)$ storage. A {\em tensor network} is a graph $G = (V,
E)$ in which each node is associated with a tensor and each edge with
an index of the adjacent tensor(s). Edges between nodes represent
shared indices that must be summed over. For example, one possible way
to represent a matrix-vector multiplication in a tensor network is
shown in Fig.~\ref{fig:tensor_network_example}. The edge $j$ between
the tensors $A_{i,j}$ and $v_{j}$ represents the operation $\sum_{j}
A_{i,j} v_j$.  A summation over shared indices is called a {\em
  contraction}. 

The tensor network associated with a quantum circuit contains a tensor
for each gate, a tensor for the initial state of each qubit, and (optionally) a
tensor for the output state under consideration on those qubit lines
for which the outcome $\bra{0}$ or $\bra{1}$ has been
specified. Edges follow the qubit lines. In this paper we consider
only single and two-qubit gates, without loss of generality since any
quantum gate can be approximated to arbitrary precision using only
single and two-qubit gates.  A single-qubit gate is represented as a
rank-2 tensor (one edge entering, one edge leaving the associated
node), corresponding to a $2 \times 2$ matrix, and a two-qubit gate is
represented as a rank-4 tensor (two edges entering, two edges
leaving), corresponding to a $4 \times 4$ matrix.

State-of-the-art algorithms for quantum circuit simulation are for the
most part based on the tensor network representation of circuits as
described above, which we will refer to as {\em circuit graphs}. A
seminal work in the area is \cite{markov2008simulating}, describing a
simulation algorithm that runs in time exponential in the treewidth of
the line graph of a circuit graph (the treewidth of the line graph is
within a constant factor of the treewidth of the corresponding circuit
graph). Numerical experiments with this methodology are presented in
\cite{fried2017qtorch}. After the first preprint of this paper was
posted on arXiv, a series of recent works
\cite{boixo2017simulation,chen201864,li2018quantum,chen2018classical,markov2018quantum,villalonga2018flexible,chen2019teleport,villalonga2019frontier,guo2019PEPS,zhang2019classical}
have tackled the problem, increasing the size or depth of circuits
that can be simulated; we will discuss the techniques used in these
papers after describing our approach.  The best known upper bounds on
the computational complexity of simulating quantum circuits of the
class studied in this paper are given in
\cite{aaronson2016complexity}.


\subsection{Simulation methodology: main building blocks}
We propose an approach that builds on the foundations of tensor
networks, with some generalizations to allow for more flexibility in
the calculations, especially when we are interested in computing the
entire state vector rather than single amplitudes. Three ideas are
crucial for our approach: (1) using hyperedges to exploit diagonal
\red(or, more in general, separable) gates, (2) allowing contractions
between non-adjacent nodes of the tensor network, and (3) tensor
slicing. The first two ideas are not explicitly discussed in the
literature on tensor networks for quantum circuit simulation, to the
best of our knowledge; slicing is used in
\cite{haner2017simulation}. We describe these ideas below.

\begin{figure}[tb]
  \begin{minipage}{0.4\textwidth}
  \leavevmode
  \centering
  \Qcircuit @C=1em @R=.7em {
    \lstick{\ket{0}} & \ustick{i_0} \qw & \gate{T} & \ustick{i_0} \qw & \ctrl{1}  & \ustick{i_0} \qw & \gate{T} & \ustick{i_0} \qw   \\
    \lstick{\ket{0}} & \ustick{i_1} \qw & \gate{H} & \ustick{j_1} \qw & \gate{Z}  & \ustick{j_1} \qw & \gate{H} & \ustick{k_1} \qw 
  }
  \end{minipage}
  \hspace*{2em}
  \begin{minipage}{0.6\textwidth}
    \includegraphics[width=0.8\textwidth]{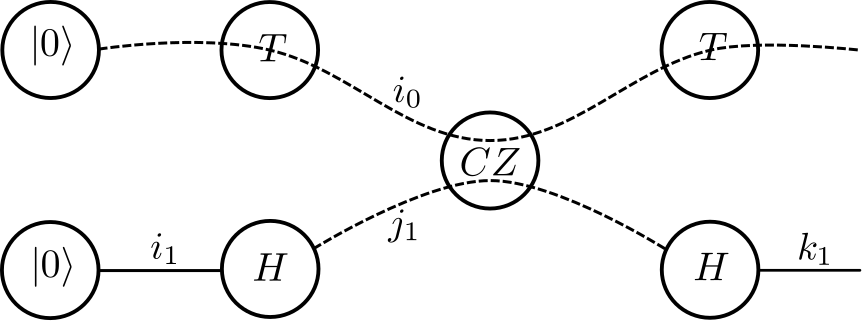}
  \end{minipage}
  \caption{Example of the proposed tensor network representation of a
    quantum circuit (left), using a hypergraph (right). The $T$ and
    $CZ$ gates are diagonal. Dashed edges are hyperedges, solid edges
    are ``regular'' edges, i.e., hyperedges of cardinality two.}
  \label{fig:hypergraph_example}
\end{figure}

\begin{figure}[tb]
  \begin{minipage}{0.4\textwidth}
  \leavevmode
  \centering
  \Qcircuit @C=1em @R=.7em {
    \lstick{\ket{0}} & \ustick{i_0} \qw & \gate{T} & \ustick{i_0} \qw & \multigate{1}{i\text{SWAP}}  & \ustick{j_0} \qw & \gate{T} & \ustick{j_0} \qw   \\
    \lstick{\ket{0}} & \ustick{i_1} \qw & \gate{H} & \ustick{j_1} \qw & \ghost{i\text{SWAP}}         & \ustick{k_1} \qw & \gate{H} & \ustick{\ell_1} \qw 
  }
  \end{minipage}
  \hspace*{2em}
  \begin{minipage}{0.6\textwidth}
    \includegraphics[width=0.8\textwidth]{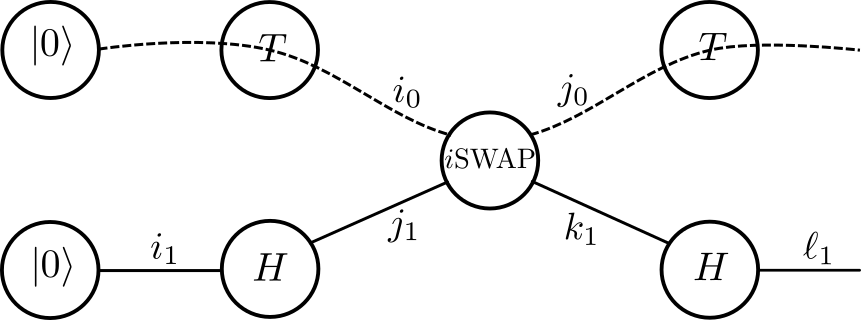}
  \end{minipage}
  \caption{Example of the proposed tensor network representation of a
    quantum circuit (left), using a hypergraph (right). The $T$ gates
    are diagonal, $i$SWAP is not \red{(although it is
      separable)}. Dashed edges are hyperedges, solid edges are
    ``regular'' edges; i.e., hyperedges of cardinality two.}
  \label{fig:hypergraph_iswap_example}
\end{figure}

Formally, we call a gate {\em diagonal} if the associated tensor
$A_{i_1,\dots,i_m,j_1,\dots,j_m}$ is nonzero only if $i_k = j_k$ for
$k=1,\dots,m$; this is a generalization of the concept of diagonal
matrices, translated to tensors. For example, the $2 \times 2$
identity matrix is a diagonal gate. \red{We call a gate {\em
    separable} if it can be obtained from a diagonal gate with a
  permutation, i.e., there exist functions $f_1,\dots,f_m$ such that
  $A_{f_1(j_1,\dots,j_m),\dots,f_m(j_1,\dots,j_m),j_1,\dots,j_m}$ is
  diagonal.} For a
diagonal gate, because all elements outside the diagonal are zero, it
is convenient to assign the same index label to input and output
edges, writing the tensor as $A_{i_1,\dots,i_m,i_1,\dots,i_m}$. This
representation is not only natural, but also more economical in that
it reduces the total number of index labels. We remark that in the
case of a sequence of diagonal gates, the same index label could be
associated with multiple edges spanning across multiple nodes of the
tensor network. Thus, rather than representing the tensor network as a
graph, we use a directed {\em hypergraph}, in which each hyperedge
consists of an ordered sequence of nodes and is associated with a
single index label. A hyperedge consisting of just two nodes is
equivalent to an edge in the original tensor network. We give examples
of this representation in Figs.~\ref{fig:hypergraph_example} and
\ref{fig:hypergraph_iswap_example}. \red{Conceptually, separable gates
  can be treated in the same way as diagonal gates: the only
  difference is that computationally we must keep track of the
  permutation whenever the tensor corresponding to the gate is
  applied. Thus, we could represent the example of
  Fig.~\ref{fig:hypergraph_iswap_example} with the same hypergraph as
  in Fig.~\ref{fig:hypergraph_example}, but any tensor operation with
  a separable, non-diagonal gate requires accounting for the
  permutation so that output indices are appropriately computed. For
  the circuits used in the large-scale simulations discussed in this
  paper, all separable two-qubit gates are also diagonal or
  diagonalizable with a straightforward basis change, hence we discuss
  diagonal gates only in the following.}

This representation of the tensor network as a hypergraph allows us to
compute tensor ranks that more accurately correspond to the real
memory occupancy in a computer implementation of the simulation
algorithm. Indeed, a rank-$k$ diagonal tensor requires only
$O(2^{k/2})$ memory, as compared to $O(2^{k})$ for non-diagonal
tensors; consequently, it is incident to $k/2$ hyperedges. This type
of consideration plays a crucial role when determining how to
partition a tensor network into manageable subgraphs (corresponding to
subcircuits). The savings become clear in constructing the line graph
of the proposed hypergraph: the line graph of a diagonal two-qubit
gate (e.g., controlled-Z gate) is a 4-clique in the traditional tensor
network representation, but is only a 2-clique in the proposed
representation. This contributes to reducing the treewidth; more
details are given in the Supplementary Information.

The second idea that is crucial for our method is the contraction of
non-adjacent nodes in the hypergraph. As mentioned earlier, existing
simulation methodologies based on tensor networks rely on determining
a sequence of contractions between adjacent vertices that leads to the
contraction of the entire graph to a single node
\cite{markov2008simulating,boixo2017simulation}. In our setting, this
is not a viable approach. Since we are interested in determining the
whole state vector for an $n$-qubit system, rather than a single
amplitude, a straightforward application of a sequential tensor
contraction method would lead to tensors of size proportional to $2^n$
simply due to the number of open ranks; i.e., the edges at the end of
the circuit that are not connected to an output state $\bra{0}$ or
$\bra{1}$. We instead allow the non-adjacent contraction of arbitrary
sets of nodes in the hypergraph. Such a contraction performs the usual
summation over shared indices (i.e., edges interior to the set being
contracted), and applies an outer product to the non-shared
indices. This approach provides additional flexibility in the order in
which contractions are performed to further reduce memory
requirements.

In broad terms, we partition the tensor network into sub-hypergraphs
corresponding to subcircuits, each of which includes fewer qubits than
the initial circuit.  Because our end goal is to compute all (or
  most) amplitudes, we perform computations within each subcircuit
following the ``Schr\"odinger approach''
\cite{aaronson2016complexity}; that is, starting from the initial
state (i.e., the generalized contraction of the tensor $\ket{0}$ for
the qubits under consideration, corresponding to an outer product), we
apply layers of quantum gates one at a time. If there are any
two-qubit gates connecting (``bridging'') two different subcircuits,
we can choose to defer one of the two corresponding (adjacent)
contractions until a subsequent stage of the computation. This
{\em contraction deferral} is accomplished via a non-adjacent
contraction that leaves open ranks in the corresponding tensors,
increasing memory consumption, but avoids merging potentially large
subcircuits. Eventually the subcircuits have to be merged
(contracted) to compute the final quantum state, or to
initialize the state of a subsequent tensor in the circuit.

\begin{figure}[tb]
  \centering
  \includegraphics[width=0.8\textwidth]{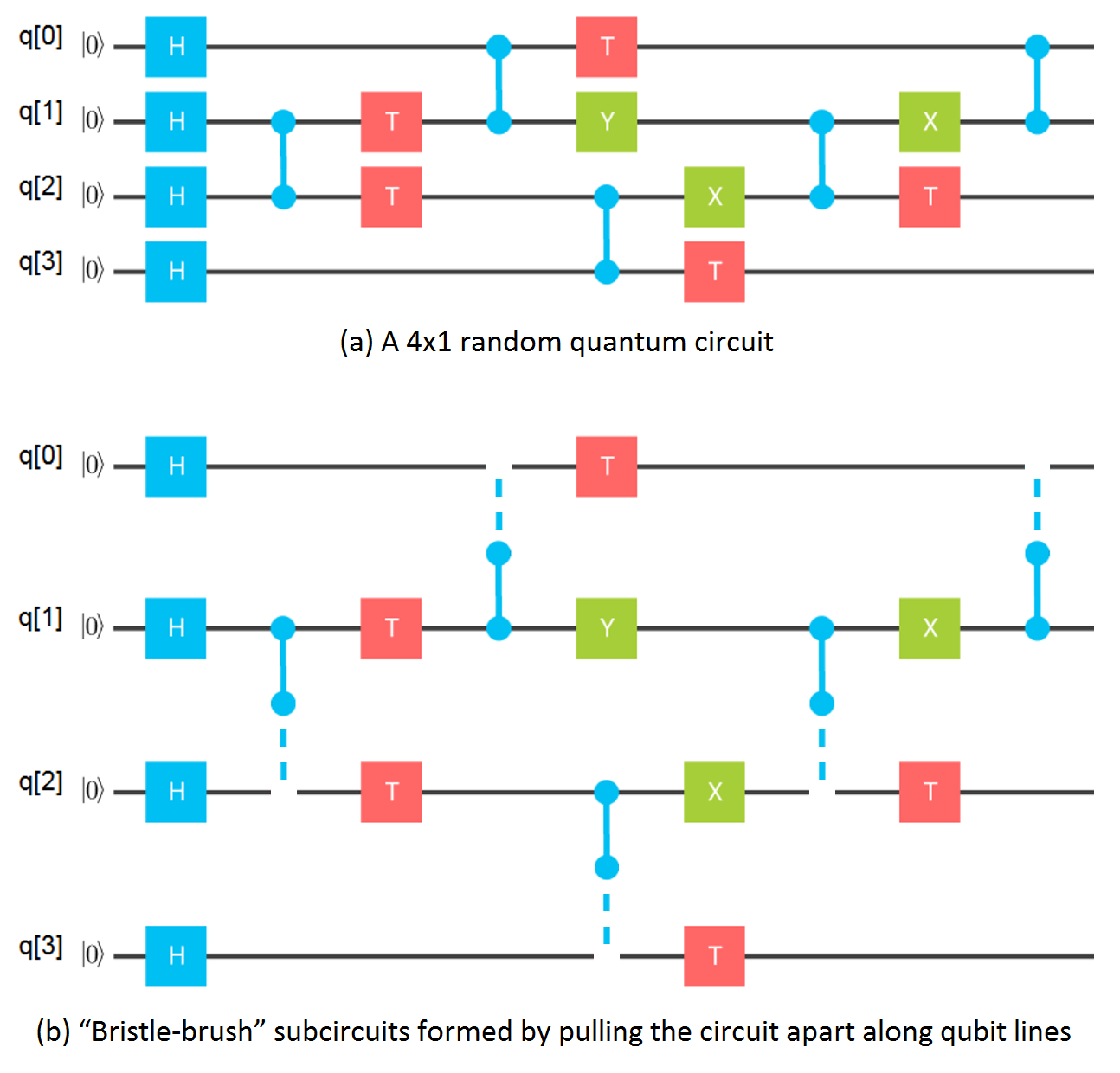}
  \caption{\red{Example from \cite{pednault2017blog} extended to
      illustrate the complete partitioning of a quantum circuit into
      ``bristle-brush'' subcircuits divided along qubit lines.
      The dashed lines correspond to {\em entanglement indices} that
      are shared between tensors constructed for each qubit line.
      Figures~\ref{fig:googleex} and~\ref{fig:googleexsplit} in the
      Supplementary Information describe alternative ways of partitioning this
      circuit that yield different computation/memory trade-offs
      during simulation.}}
  \label{fig:bristlebrush}
\end{figure}

\begin{figure}[tb]
  \centering
  \begin{minipage}{0.45\textwidth}
    \centering
    \includegraphics[width=0.58\textwidth]{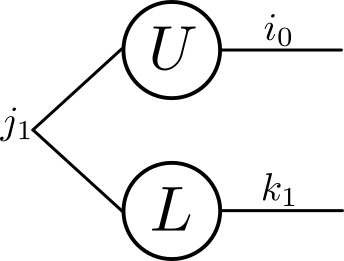}
  \end{minipage}
  \hfill
  \begin{minipage}{0.45\textwidth}
    \centering
    \includegraphics[width=0.7\textwidth]{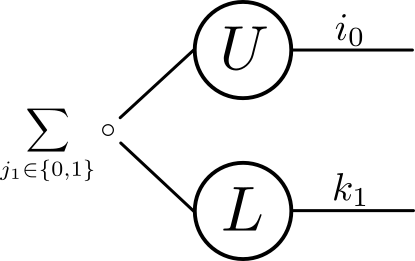}
  \end{minipage}
  \caption{{\em (a) Left.} An example of contraction deferral:
      the contraction operation on index $j_1$ in the tensor network
      of Fig.~\ref{fig:hypergraph_example} is deferred.  Standard
      adjacent contractions are performed to construct the upper
      tensor $U_{i_0,j_1} = T_{i_0} CZ_{i_0j_1,i_0j_1} T_{i_0}
      \delta_{i_0}$ (we denote each tensor by the name of the
      corresponding gate, and denote the single-index Kronecker delta
      by $\delta_k$).  A non-adjacent contraction is performed on the
      Hadamard gates to construct the lower tensor $L_{k_1,j_1} =
      H_{k_1,j_1} \sum_{i_1 \in \{0,1\}}{H_{j_1,i_1} \delta_{i_1}}$.
      Contraction deferral allows the top and bottom tensors,
      $U_{i_0,j_1}$ and $L_{k_1,j_1}$, to be computed independently.
      The final state is obtained by contracting $j_1$.  We refer to
      indices whose contractions are deferred as {\em entanglement
      indices}, in recognition of the fact that they account for the
      entanglements that exist among subcircuits while allowing those
      subcircuits to be simulated independently.  In this example,
      $j_1$ is an entanglement index. {\em (b) Right.} An example of
      sliced contraction deferral: a deferred contraction is combined
      with the slicing of entanglement index $j_1$. For each $j_1 \in
      \{0,1\}$, the top and bottom tensors, $U_{i_0,j_1}$ and
      $L_{k_1,j_1}$, are computed independently and their values are
      multiplied together.  The resulting products are then summed
      over the values of $j_1$.  The contraction operation on $j_1$ is
      thus accomplished iteratively.  Because $j_1$ is fixed at each
      iteration, the tensors $U_{i_0,j_1}$ and $L_{k_1,j_1}$ are
      sliced on $j_1$ and the amount of memory needed to store the
      slices is cut in half.}
  \label{fig:deferral_example}
\end{figure}

\begin{figure}[tb]
  \centering
  \begin{minipage}{0.45\textwidth}
    \centering
    \includegraphics[width=0.58\textwidth]{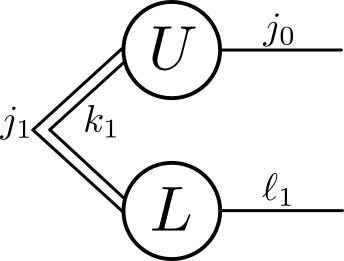}
  \end{minipage}
  \hfill
  \begin{minipage}{0.45\textwidth}
    \centering
    \includegraphics[width=0.7\textwidth]{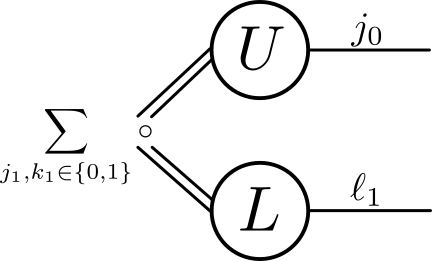}
  \end{minipage}
  \caption{{\em (a) Left.} An example of contraction deferral: the
    contraction operation on indices $j_1,k_1$ in the tensor network
    of Fig.~\ref{fig:hypergraph_iswap_example} is deferred.  Standard
    adjacent contractions are performed to construct the upper tensor
    $U_{j_0,j_1k_1} = T_{j_0,j_0} \sum_{i_0 \in \{0,1\}}
    i\text{SWAP}_{j_0k_1,i_0j_1} T_{i_0,i_0} \delta_{i_0}$.  A
    non-adjacent contraction is performed on the Hadamard gates to
    construct the lower tensor $L_{\ell_1,j_1k_1} = H_{\ell_1,k_1}
    \sum_{i_1 \in \{0,1\}}{H_{j_1,i_1} \delta_{i_1}}$.  Contraction
    deferral allows the top and bottom tensors, $U_{j_0,j_1k_1}$ and
    $L_{\ell_1,j_1k_1}$, to be computed independently.  The final
    state is obtained by contracting $j_1$ and $k_1$. In this example,
    $j_1$ and $k_1$ are entanglement indices. {\em (b) Right.}  An
    example of sliced contraction deferral: a deferred contraction is
    combined with the slicing of entanglement indices $j_1, k_1$. For
    each $j_1,k_1 \in \{0,1\}$, the top and bottom tensors,
    $U_{j_0,j_1k_1}$ and $L_{\ell_1,j_1k_1}$, are computed
    independently and their values are multiplied together. The
    resulting products are then summed over the values of $j_1$ and
    $k_1$.  Because $j_1$ and $k_1$ are fixed at each iteration, the
    tensors $U_{j_0,j_1k_1}$ and $L_{\ell_1,j_1k_1}$ are sliced on
    $j_1,k_1$ and the amount of memory needed to store the slices is
    cut by a factor of four.}
  \label{fig:deferral_iswap_example}
\end{figure}

The simulation strategy described in \cite{pednault2017blog} provided
the original motivation for contraction deferral.  This strategy
involves {\em ``pulling a grid circuit apart into individual `bristle
  brushes,' one for each qubit, then computing the corresponding
  tensors, and finally combining the tensors for each qubit to
  calculate the quantum amplitudes for the overall circuit.''}  An
example is shown in Fig.~\ref{fig:bristlebrush} \footnote{The name
  ``bristle brushes'' comes from \cite{pednault2017blog}; see also
  Fig.~3 of \cite{villalonga2018flexible} and
  \cite{villalonga2019frontier}.}.  If we take the individual qubit
lines to correspond to subcircuits, tensors corresponding to
entangling gates must be assigned to exactly one of the subcircuits
that they bridge. The choice can be made arbitrarily but it can have
computational implications. The construction of per-qubit tensors
involves non-adjacent contractions of gate tensors to effectively jump
over entangling gates that were assigned to other qubit lines, as
indicated with dashed lines in Fig.~\ref{fig:bristlebrush}.
An example of contraction deferral is given in
Figs.~\ref{fig:deferral_example}{\it(a)} and
\ref{fig:deferral_iswap_example}{\it(a)}.  As illustrated, contraction
deferral introduces {\em entanglement indices} to account for
yet-to-be-resolved entanglements among subcircuits.

To the best of our knowledge, the first paper on quantum circuit
simulation to use a form of contraction deferral is
\cite{aaronson2016complexity}. In fact, the use of contraction
deferral is not limited to the strategy discussed above, and is
independent of the order of contraction. We demonstrate this fact in
the Supplementary Information by showing how circuit decompositions
based on \cite{aaronson2016complexity}, together with a bidirectional
contraction order, can in principle be used to calculate single
amplitudes of $7 \times 7$, depth 46 circuits using 141~TB of memory
in a matter of a few hours on Vulcan-class supercomputers.  In that
case, the contractions proceed from both the initial and final states
inward, toward the middle, and contraction deferral is employed in
both directions.

The third and final building block of our methodology is tensor
slicing. The idea is to select certain hyperedges in the hypergraph,
say $s$ of them, loop over the $2^s$ possible combinations of values
for these hyperedges, and perform the remaining computations with the
value for these hyperedges fixed. This yields $2^s$ ``sliced''
tensors. If the choice is made appropriately, the resource consumption
of each sliced tensors is a factor $2^s$ smaller than the whole
circuit. Since there are $2^s$ sliced tensors, it may seem that this
does not yield any savings; however, our approach tries to reorder the
tensor computations so that only a few selected slices (rather than
the full tensor) need reside in primary storage, thus decreasing
resource consumption and better exploiting parallelism. We also
utilize slicing to make the use of secondary storage viable by
minimizing data transfer between primary and secondary storage. Our
scheme extends the simulation strategy in \cite{haner2017simulation}:
we choose a set of qubits to slice, looping over every possible
combination of their values, and a superset of those qubits to
efficiently index secondary storage. Depending on whether or not
slicing is applied to a hyperedge whose contraction is deferred, we
distinguish between {\em sliced contraction deferral} and regular
(i.e., non-sliced) contraction
deferral. Figs.~\ref{fig:deferral_example}{\it(b)} and
\ref{fig:deferral_iswap_example}{\it(b)} provide examples of sliced
contraction deferral applied to the tensor networks of
Figs.~\ref{fig:hypergraph_example} and
\ref{fig:hypergraph_iswap_example}. It should be noted that, while
sliced contraction deferral reduces the memory footprints needed to
calculate subcircuit tensors, it does not reduce the memory
requirements for storing and summing the contraction results, nor does
it reduce the computational requirements. While we did not use sliced
contraction deferral in our simulations, it is a key component of the
recent literature (see next section), hence it is important to
highlight it. The mathematical details of our approach are provided
in the Supplementary Information, where we formalize our methodology
and use the circuit shown in Fig.~\ref{fig:bristlebrush} as a running
example to illustrate further ideas.

\begin{figure}[tb]
  \centering
  \includegraphics[width=0.75\textwidth]{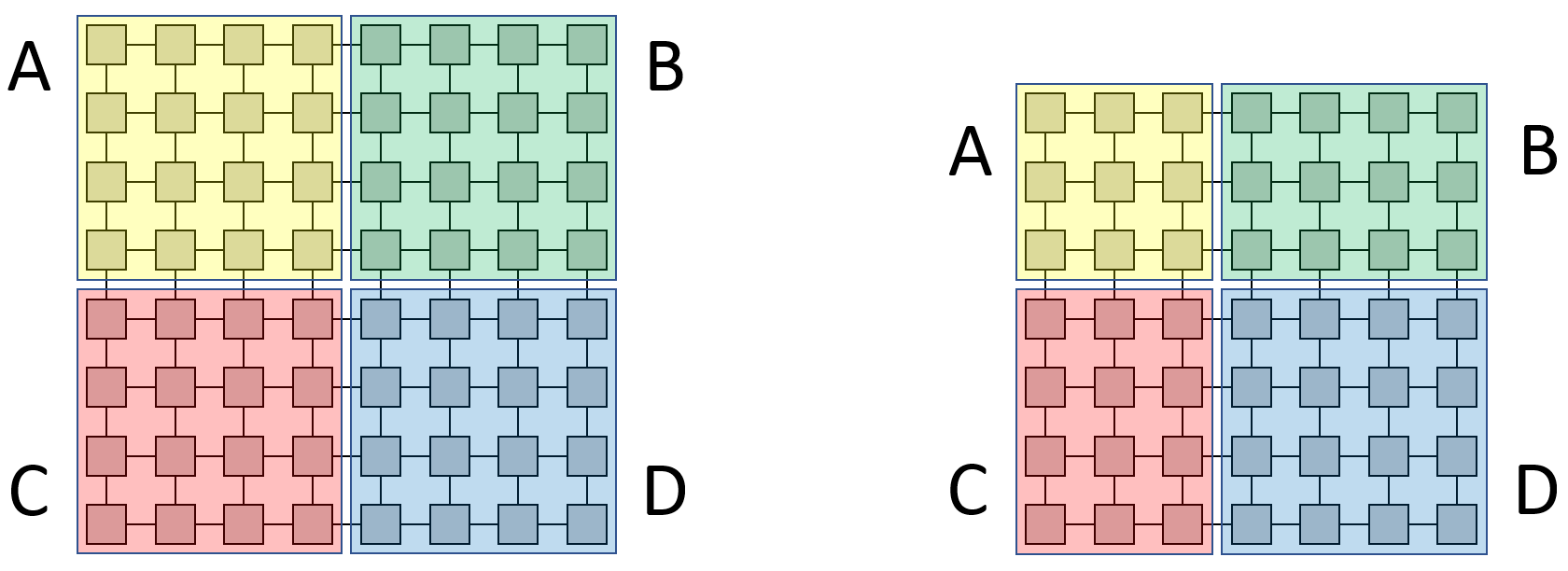}
  \caption{{\em (a) Left.} Partitioning of an $8 \times 8$-qubit
      circuit into four $4 \times 4$-qubit subcircuits when the
      circuit is viewed along the qubit lines (i.e., down the axes of
      the ``bristle brushes''). {\em (b) Right.}  Corresponding
      partitioning for a $7 \times 7$-qubit circuit.}
  \label{fig:bristlepartition}
\end{figure}

We now provide several examples of slicing to illustrate the
trade-offs involved that affect overall performance. Consider the $8
\times 8 \times (1 + 8 + 1)$ circuit in the GRSC repository
\cite{markov2018quantum,boixo2018repository}, which is a $8 \times
8$-qubit, depth 8 circuit with an additional layer of Hadamard gates
at the start of the circuit and another at the end of the circuit. We
can simulate this circuit as follows. We first construct subcircuits
(tensors) for the individual qubit lines, using the ``bristle-brush''
strategy. Per \cite{pednault2017blog}, we then contract these tensors
into four $4 \times 4$ subcircuits as indicated in
Fig.~\ref{fig:bristlepartition}{\em (a)}. Each resulting subcircuit
tensor has 16 output indices and 8 entanglement indices, yielding
double-precision memory footprints of $2^{16 + 1\cdot 8 + 4} = 256$~MB
each. We finally calculate the individual output amplitudes by setting
the values of the output indices according to a desired amplitude,
thereby slicing the four intermediate tensors; this yields four tensor
slices of size $2^{1\cdot 8 + 4} = 4$~KB each, which are contracted to
yield the desired amplitude.

In general, it is desirable to perform tensor slicing operations as
early in a simulation as possible, because doing so can reduce memory
requirements without necessarily increasing the number of
floating-point operations that are performed per amplitude calculated.
In the above example, we can slice the output index of the initial
tensors corresponding to the qubit lines {\em before} contracting them
into the $4 \times 4$ subcircuits. This early slicing approach reduces
the maximum memory footprints of the $4 \times 4$ subcircuits from
256~MB to 4~KB each.

If we apply the same strategy to the $8 \times 8 \times (1 + 40 + 1)$
circuits in the GRSC repository (i.e., 64-qubit, depth 40 circuits
with initial and final layers of Hadamard gates), the corresponding
four subcircuits require $2^{5 \cdot 8 + 4} = 16$~TB of memory each,
assuming early slicing of the output indices as described
above. Contracting these four tensors in pairs (i.e., A with B,
followed by C with D, followed by AB with CD) requires an additional
16~TB of temporary memory to hold intermediate calculations, resulting
in a total memory footprint of 80 TB for this final stage of
calculation.  In the case of $7 \times 7 \times (1 + 40 + 1)$ circuits
paritioned as shown in Fig.~\ref{fig:bristlepartition}{\em (b)},
subcircuit D requires 16~TB while each other subcircuit requires at
most 0.5~TB.  The temporary space needed to perform the final
contractions is 0.5~TB, yielding a total memory footprint of 18~TB for
the final stage of calculation.

Although such memory footprints fit comfortably within the resources
available on current classical computers, the computational demands
entailed by the above simulation strategy are quite high.  For $7
\times 7 \times (1 + 40 + 1)$ circuits, the first two of the final
contractions (i.e., tensors A with B and C with D) require $2^{50}$
and $2^{55}$ complex multiplications and additions per amplitude,
while the final contraction requires $2^{35}$.  Taken together, this
translates to approximately $2^{58}$ floating-point operations per
amplitude.  For $8 \times 8 \times (1 + 40 + 1)$ circuits, the three
final contractions together require approximately $2^{61}$ complex
multiplications and additions per amplitude, or $2^{64}$
floating-point operations per amplitude. In the case of the $7 \times
7$ depth 27 circuit whose simulation we present later, the above
strategy requires approximately $2^{40}$ floating-point operations per
amplitude for the final three contractions, using a memory footprint
of only 4.9 GB. However, to compute all $2^{49}$ amplitudes, these
contractions would require at least 4,168 years to complete at the
5~PFlops/sec peak performance of the Vulcan supercomputer employed in
our simulations, which is clearly infeasible. To achieve our objective
of computing all amplitudes, we therefore optimize the circuit
partitioning at the expense of increased memory requirements to make
more efficient use of contraction deferral and significantly speed up
the calculations, reducing the number of floating-point operations per
amplitude to $2,916 \approx 2^{11.5}$, a factor of $\approx 2^{28.5}$
times less computation.  The optimization methodology is discussed in
a subsequent section.

\subsection{Tensor contraction deferral in the literature}
Before turning our attention to the question of optimization, we
review the recent literature on quantum circuit simulation methods
based on tensor networks, to provide a unifying perspective showcasing
the importance of contraction deferral and slicing, also when combined
together.

Sliced contraction deferral, or an equivalent methodology, is used in
\cite{chen201864,li2018quantum,chen2018classical,markov2018quantum,villalonga2018flexible,villalonga2019frontier,zhang2019classical}
to reduce both memory requirements and interprocessor communication.
In a majority of these methods, the reductions achieved enable all
intermediate tensor slice computations to fit within the memories of
individual processing nodes for the circuits simulated, thereby
eliminating the need for interprocessor communication except to
perform final summations of the resulting tensor products.  This
approach has the benefit of maximizing the degree of parallelism that
can be achieved while simultaneously eliminating communication
overhead, which in some applications can be orders of magnitude
greater than that of the computations.  However, small memory
footprints (and accompanying low communication requirements) can come
at the cost of very high computational requirements, as the
``bristle-brush'' strategy demonstrates; hence, such approaches do not
guarantee superior performance.

In \cite{villalonga2018flexible,villalonga2019frontier}, sliced
contraction deferral is applied in combination with non-sliced
contraction deferral. Circuits are first contracted along qubit lines
to produce grids of per-qubit tensors; these per-qubit tensors
correspond to the example in Fig.~\ref{fig:bristlebrush}, and their
construction requires contraction deferral. Grids of per-qubit tensors
are then ``cut'' \cite{villalonga2018flexible} into subcircuits using
``systematic tensor slicing'' \cite{villalonga2019frontier} (i.e.,
sliced contraction deferral), thereby enabling the subcircuits to be
contracted independently before the final merge; this subcircuit
decomposition corresponds to the approach outlined in
\cite{pednault2017blog}. Because these subcircuits share a large
number of entanglement indices for deep circuits, the use of sliced
contraction deferral leads to a significant reduction in both memory
requirements and interprocessor communication.

The approach presented in \cite{chen2018classical,zhang2019classical}
uses a ``variable fixing'' operation that is a contraction slicing
method. It is applied both across qubit lines to entangling gates, as
discussed in this paper, and to along-qubit-line indices as in the
depth-wise slicing strategy of \cite{aaronson2016complexity}.
\cite{chen2018classical,zhang2019classical} uses an optimization
algorithm to determine which contractions should be deferred through
slicing.  Contractions that are not sliced are performed using an
extension of the approach in \cite{markov2008simulating} generalized
to efficiently handle diagonal gates. Because
\cite{markov2008simulating} and
\cite{chen2018classical,zhang2019classical} only consider adjacent
contractions, they explore different circuit decompositions from what
is proposed in this paper. The inclusion of depth-wise slicing first
introduced in \cite{aaronson2016complexity} is also a unique
characteristic of this approach, relative to other tensor-network
approaches.

Taken together, the approaches presented in
\cite{villalonga2018flexible,villalonga2019frontier} and in
\cite{chen2018classical,zhang2019classical} have been used to simulate
some of the largest benchmark circuits considered to date.

Interestingly, sliced contraction deferral across qubit lines is not
directly considered in \cite{aaronson2016complexity}.  Instead,
\cite{aaronson2016complexity} introduces a gate decomposition approach
that is also employed --- with some variations --- in
\cite{chen201864,li2018quantum,markov2018quantum}.  As mentioned in
\cite{markov2018quantum}, gate decompostion can be understood in terms
of the Schmidt decomposition of a matrix. In the context of $4 \times
4$ matrices, for any matrix $M \in \mathbb{C}^{4 \times 4}$, there
exist matrices $U_k,V_k \in \mathbb{C}^{2 \times 2}$ and nonnegative
reals $s_k \in \mathbb{R}_{+}$, $k=1,\dots,4$, such that $M =
\sum_{k=1}^4 s_k U_k \otimes V_k$.  This equation can be obtained from
the singular value decomposition of (a permutation of) $M$.  Notice
that the Schmidt rank could be strictly less than four; i.e., some
$s_k$ could be zero. Thus, a two-qubit gates can be decomposed into a
sum of tensor products of single-qubit (not necessarily unitary)
matrices. The decomposition can be substituted directly in the tensor
equations that define a circuit, enabling subcircuits to be simulated
independently without increasing memory footprints, just as in sliced
contraction deferral. We remark that gate decomposition in its most
general form is different from sliced contraction deferral.  However,
the gate decompositions considered in
\cite{chen201864,li2018quantum,markov2018quantum} are a special case
that can be seen as equivalent. Indeed, all three methods employ the
$CZ$-gate decomposition:
\begin{equation}
  CZ = \ket{0}\bra{0} \otimes I + \ket{1}\bra{1} \otimes Z.
\end{equation}
This is also a form of sliced contraction deferral. By rewriting the
gate decomposition in tensor form $CZ_{jk,jk} = \delta_{j,0} +
\delta_{j,1}Z_{k,k}$, where $\delta_{m,n}$ is the two-index Kronecker
delta, it becomes evident that the index $j$ is effectively sliced by
the Kronecker deltas and its contraction is performed at the end of
the computation, as indicated in \cite{li2018quantum}. The gate
decomposition is instead used in its most general form within the
Projected Entangled Pair States (PEPS) approach presented in
\cite{guo2019PEPS}; this approach computes a PEPS for the final
quantum state by directly applying the single-qubit matrices obtained
from the singular value decomposition of the two-qubit gates. A tensor
network computation is still necessary to perform a projective
measurement on the final PEPS, but it does not have a precise
correspondence to sliced contraction deferral.

To the best of our knowledge, there is no single implementation that
tries to combine all of the circuit-decomposition methods discussed above,
and it seems clear that an efficient optimization algorithm to
determine the best strategy, within the large space of all
possibilities, could have a large impact on practical performance.

\subsection{Optimizing the circuit simulation strategy}
In general, due to the large number of possible circuit
decompositions, several considerations need to be taken into account
to determine an appropriate simulation execution strategy. These
considerations will be discussed next.

For a given input quantum circuit, there is an exponentially large
number of possible ways of partitioning circuits and sequencing the
computations. Additionally, for every non-diagonal entangling gate
whose contraction could potentially be deferred, we have the choice of
keeping it as is, or transforming it through circuit rewriting to
either: diagonalize the gate so as to reduce the number of
entanglement indices that would be introduced (e.g., such as
transforming a $CX$ gate into two Hadamard gates and one $CZ$ gate);
or apply gate decomposition in cases where the minimum number of terms
in the decomposition (the Schmidt rank of the gate) is less than the
size of the contraction summation induced by the corresponding
entanglement indices (i.e., 4 in the case of non-diagonal two-qubit
gates).  These options generate a space of possible computation
schemes for any given circuit. Given a circuit and a computation
scheme, we can compute the memory and floating-point operations
required simply by analyzing subcircuits. In principle, we would like
to run a multi-objective optimization over the space of possible
computation schemes to generate the Pareto frontier that defines the
optimal trade-off between memory usage and floating-point operations.
It is desirable to have an automated process to determine an optimal
computation scheme, and it is clear that a brute force exploration
strategy of all combinations is not viable for circuits above a modest
size.

To obtain the numerical results presented in this paper, we employ
two heuristic search strategies: one that implements some
hand-crafted circuit partitioning criteria based on our intuitions,
and one that explores the choices described above in a largely
depth-first manner. The hand-crafted strategies adopt the following
principles:
\begin{itemize}
\item Diagonalize $CX$ gates if contraction deferral is being applied, or
  if all subsequent gates applied to the
  control qubit are diagonal or diagonalizable.
\item Try to keep each circuit partitioned into a small number of
  (e.g., less than five) subcircuits, with a small number of gates
  crossing the boundaries.
\end{itemize}
The Supplementary Information illustrates how the 49-qubit, depth 27
random circuit and the 56-qubit, depth 23 random circuit used in our
numerical study are partitioned into subcircuits using the
hand-crafted approach described above.

We also use a largely depth-first heuristic search to explore the
trade-off between memory usage and floating-point operations for
various circuits. The method works as follows. It is given a circuit
and an upper bound to the memory consumption allowed for the
simulation of that circuit. Iteratively, the heuristic selects a
two-qubit gate that bridges a pair of subcircuits that have not yet
been merged. It then creates two decision branches: one in which the
two subcircuits are merged by performing a contraction, the other one
in which the contraction is deferred.  A decision branch can be pruned
if it results in exceeding the memory consumption allowed. The search
space is explored using $A^{*}$ \cite{astar}, where the heuristic
function to assign node potentials is given by a lower bound on the
total number of floating point operations resulting from the remaining
decisions. Since the lower bound employed is quite weak, we maintain
multiple $A^{*}$ search queues representing different ranges of search
depths. We process the search queues in a probabilistic round-robin
fashion that favors greater search depths, resulting in a largely
depth-first search guided by the $A^{*}$ heuristic.  A lower bound on
the number of remaining floating-point operations can be readily
obtained by calculating the costs of applying any remaining gates
assuming that the existing tensors will not increase in size; i.e., by
treating any remaining entangling gates that bridge existing
subcircuits as if they do not bridge them and instead lie solely
within one of the subcircuits being bridged.

We note that the above, largely depth-first, search algorithm is the
same as the one referenced in our early preprint
\cite{pednault2017breaking}, and as such it considers only non-sliced
contraction deferral across qubit lines.  The results presented here
using the above algorithm thus serve as baselines that illustrate the
extent to which non-sliced contraction deferral across qubit lines
contributes to resource-efficient quantum circuit simulation.  The
memory footprints reported take into account all primary tensor data
structures that would need to be allocated and de-allocated during the
course of simulations, but they exclude the communication buffers
typically present in distributed implementations.  The reported
floating-point operations per gate per amplitude are determined by
simply counting the float-point operations (FLOPs) that would need to
be performed for the corresponding circuit partitionings.  These
averages consider the case in which all amplitudes are to be
calculated.  Communication costs are not included in the assessment.

\begin{figure}[tb]
  \centering
  \includegraphics[width=0.75\textwidth]{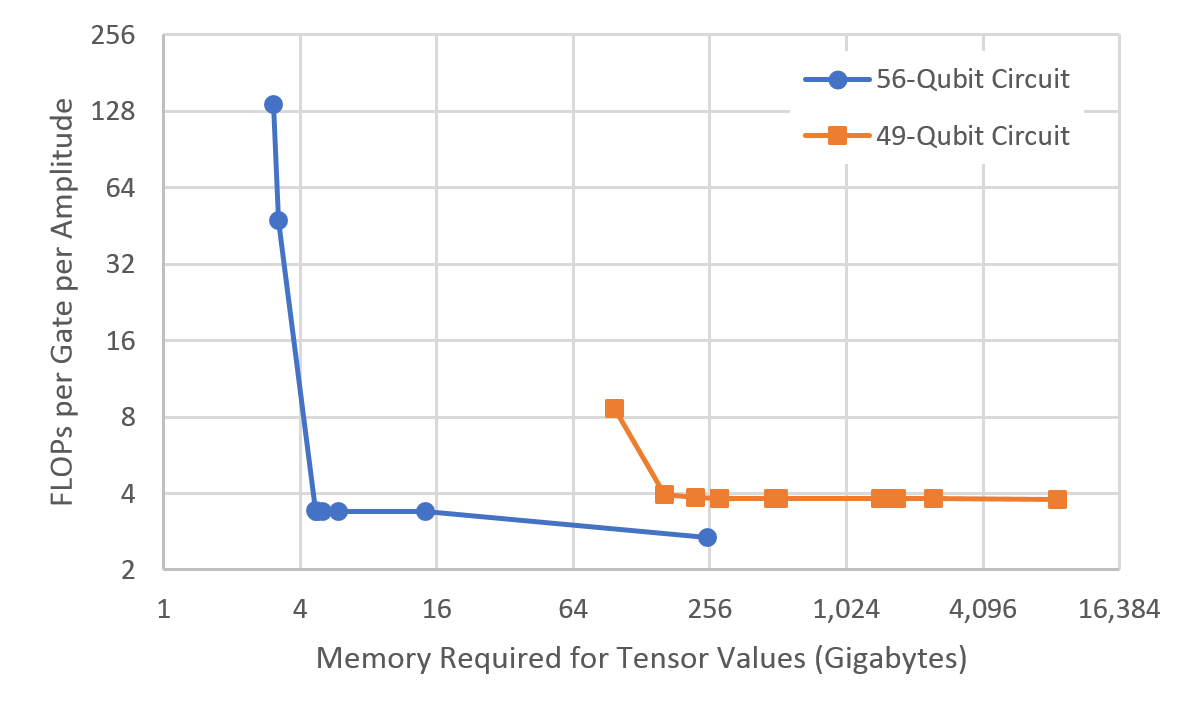}
  \caption{Trade-off between floating point operations and memory usage
    when employing only non-sliced contraction deferral across qubit lines.}
  \label{fig:tradeoff}
\end{figure}

Fig.~\ref{fig:tradeoff} illustrates the trade-off between memory and
computation identified by this analysis for the 49-qubit, depth 27 and
56-qubit, depth 23 circuits presented in the Supplementary Information
that were used in our simulation study. For both circuits, the
trade-offs reach the lowest FLOPs (per gate per amplitude) with
relatively low memory requirements (i.e., 3.78~FLOPs at 8.5~TB for the
49-qubit circuit and 2.69~FLOPs at 249~GB for the 56-qubit circuit),
and we could not improve upon these values even when larger tensors
were allowed.  Remarkably, these curves remain relatively flat as the
allowed memory footprint is decreased, reaching only 3.95~FLOPs per
gate per amplitude with a 162~GB memory allowance for the 49-qubit
circuit and 3.42~FLOPs with a 4.6~GB memory allowance for the 56-qubit
circuit.  By comparison, a complex multiply requires six
floating-point operations: four multiplications and two additions.
For both circuits, computational costs rise sharply when memory
thresholds are decreased below these levels, producing ``bends'' in
the curves.  Because smaller memory footprints imply less
interprocessor communication and greater opportunity for parallelism,
some of the simulation strategies identified in in
Fig.~\ref{fig:tradeoff} would likely lead to significantly lower
execution times for the simulations as compared to the strategies that
we actually ran.  Note also that, since the $A^{*}$ search strategy
employed was given a time limit, it is possible that this curve is
suboptimal and circuit partitioning schemes that strictly dominate
those in Fig.~\ref{fig:tradeoff} could potentially be found.  In
addition, the search complexity for our $A^{*}$ implementation
increases as the memory threshold is decreased, owing to the fact that
the lower bound on remaining computation incorporated into our $A^{*}$
heuristic becomes weaker for smaller memory thresholds.  As a result,
the time limit prevents additional circuit partitionings to be
identified below the memory levels indicated in
Fig.~\ref{fig:tradeoff}.  We note, however, that the memory-efficient
``bristle-brush'' strategy lies within the search space considered by
the $A^{*}$ algorithm, so in principle these memory thresholds could
be lowered much further given sufficient search time.
    
Complementing the $A^{*}$ approach outlined above, we develop an
integer programming model to search for an optimal circuit
partitioning. The model is based on the hypergraph representation
discussed above, and it can be solved with existing integer
programming software. In the Supplementary Information, we describe
the full model and report on our computational experience using the
IBM ILOG CPLEX optimization package to solve the instances of the
integer programming model corresponding to the two circuits discussed
above. Although we cannot solve these problem instances to global
optimality within a few hours, the software finds partitioning schemes
very similar to the one that we used, and we are able to report bounds
on the quality of these partitionings. The $A^{*}$ approach and the
integer programming model can be used to automatically determine an
efficient partitioning for a given circuit.

The methodologies presented above concern the determination of a
circuit partitioning. When secondary storage is used, it is crucial to
organize data on disk in an effective way, which in general differs
from how it is organized in primary storage. This is because data
organization on secondary storage must remain static throughout a
simulation. The Supplementary Information presents a greedy optimization method for
simultaneously choosing circuit partitionings and data layouts on
secondary storage to reduce overall cost.


\subsection{Software and hardware platform}
We implement the in-memory quantum circuit simulation approach described above
using the Cyclops Tensor Framework (CTF)~\cite{solomonik2014massively}, a
distributed-memory C++ library for tensor summations and
contractions. The library provides a domain-specific language for
tensor operations using an Einstein-summation syntax.  CTF tensors are
distributed over all processors with tensor summations and
contractions performed in a data-parallel manner using
MPI~\cite{gropp1999using}, OpenMP~\cite{dagum1998openmp}, and
BLAS~\cite{lawson1979basic}.


The massively-parallel tensor-contraction calculations enabled by CTF
have heretofore been driven by applications in computational chemistry
and physics~\cite{Bartlett:1981:ARPC:CC,orus2014practical} that
involve contractions of tensors of rank 4--8; the quantum circuit
models employed in our calculation use tensors of much higher rank.
In performing these contractions using CTF, the main challenge is the
need for higher-dimensional virtual topologies (i.e., the
decomposition of tensors among more dimensions) than CTF typically
performs.  CTF operates by mapping tensor dimensions onto dimensions
of a processor grid, often redistributing tensors to new mappings at
contraction time.  As each index in our tensor corresponds to a
  qubit in the underlying circuit, mappings to high-dimensional
  virtual processor grids are necessary.  To achieve these, we
increase the space of virtual topologies CTF considers, 
improve tthe mapping logic in CTF, and enable dynamic creation
and destruction of MPI communicators (defined for each processor grid
dimension).

We use the built-in profiling capabilities of CTF and the
performance-counter libraries made available on the hardware platform we
use in our experiments to determine the bottlenecks CTF encounters
during circuit simulation contractions \cite{WalkupPerf}.  For reasons
of brevity, we do not describe all optimizations in this paper.  Local
copy, summation, and multiplication primitives are automatically
replaced by appropriate optimized variants, chosen with knowledge of
the parameters in use during the circuit simulations.  These
low-level changes substantially improve the performance with respect
to the original implementation.

Our experiments are executed on Vulcan, a 24,576-node IBM Blue Gene/Q
supercomputer \cite{BGchip2013}; a brief description of this
  architecture follows. A Blue Gene/Q node consists of 18 A2 PowerPC
64 bit cores, 16 of which are application-accessible. Each A2 core
runs at 1.6 GHz, has a 16~KB private level 1 (L1) cache, as well as a
2~KB prefetching buffer (L1P). All the cores on the same processor
share a 32~MB level 2 (L2) cache and 16 GB of main memory. The compute
nodes are connected via a 5D torus network with a total network
bandwidth of 40 GB/s.

\section{Results}
We now present the results of our experiments on Vulcan, as well as a
numerical analysis of the sources of the difficulty of the circuits
described in \cite{markov2018quantum,boixo2018repository}.

\subsection{Simulation of a 49-qubit and a 56-qubit circuit on Vulcan}
In the case of the
49-qubit circuit, the final quantum state is calculated in $2^{11}$
slices with $2^{38}$ amplitudes each. In the case of the 56-qubit
circuit, only one slice with $2^{37}$ amplitudes is calculated (out of
the $2^{19}$ such slices defining the final quantum state).

All experimental results are obtained over the course of two days.
This time includes the time to set up the
experiments described in this paper, as well as that to conduct
additional experiments beyond the scope of what is described here.

Memory usage requires scaling to 4,096 nodes, with 64~TB of aggregate memory,
for each (parallel) calculation. The theoretical minimum memory
footprint to hold a slice of the state vector is less than 5~TB for
the 49-qubit circuit ($2^{38}$ complex doubles are used to store the
$2^{38}$ amplitudes); it is common in the literature to report only
the memory required for the state vector, e.g.,
\cite{haner2017simulation}. In practice, experiments require more
than 32~TB. The bulk of these operations require two large tensors to
compute a result and CTF internal buffering introduces approximately a
factor of 4 in overhead. Memory is also needed to reformat
contraction operands for efficiency and for buffer space as messages
are passed between nodes.

After computing the amplitudes, we analyze the distributions of the
corresponding outcome probabilities to verify whether the observed
distributions of probabilities adhere to the theoretically expected
truncated exponential (Porter-Thomas) distributions
\cite{boixo2018supremacy}.  Because outcome probabilities vary over
several orders of magnitude and scale inversely with the number of
outcomes, histograms are accumulated for log-transformed outcome
probabilities $z=\log(Np)$, where $N=2^n$, $n$ being the number of
qubits and $p$ being an outcome probability.  This mapping yields the
expected truncated Gumbel distribution for the resulting histogram:
\begin{equation}
  \label{eq:gumbel}
  f_z(z) =\left\{
  \begin{array}{ll}
  \frac{e^{z-e^z}}{1-e^{-N}} & z \leq
  n\ln{2} \\ 0 & \textrm{otherwise}.\\
  \end{array}
  \right.
\end{equation}

\begin{figure}[tb]
  \centering
  \includegraphics[width=0.8\textwidth]{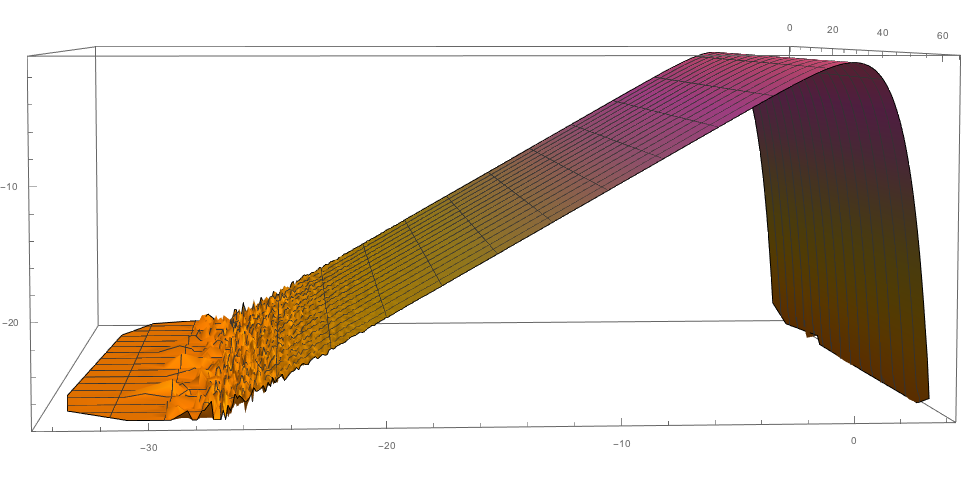}
  \caption{Histograms of log-transformed outcome probabilities for the
    individual slices of the 49-qubit circuit rendered together as a
    surface plot. The vertical axis is plotted on a log-scale.}
  \label{fig:amplitudeplot}
\end{figure}

Fig.~\ref{fig:amplitudeplot} shows a plot of the histograms we obtain
for an aggregation of the slices calculated for the 49-qubit circuit.
Each histogram is an aggregation of 32 slices, and the 64 individual
histograms are stacked together and rendered as a surface plot
to better reveal variations between the histograms.  As can be seen in
this figure, the variations are concentrated at the extreme tail ends
of the distribution where very low probabilities and, hence, very low
bin counts are observed in the histograms. Such variations are
expected in histograms whenever one is dealing with low bin counts.
Fig.~\ref{fig:amplitudeplot} reveals that the distributions of outcome
probabilities across slices are virtually identical, demonstrating
that this distribution is essentially isotropic across the quantum
state, as expected, and there is an extremely
close agreement between the observed and predicted distributions.

In the Supplementary Information, we discuss further quantum circuit
simulations enabled by the ideas presented in this paper that involve
circuits with more layers of gates. While we do not carry out these
simulations, our framework allows us to verify that the memory and
floating point operation requirements for these computations are
viable on Vulcan or similar supercomputers. More specifically,
combining targeted slicing with sporadic disk read/write operations,
as already suggested by \cite{haner2017simulation}, we provide a
method for simulating $7 \times 7$ universal random circuits up to
arbitrary depth, writing all amplitudes to disk. Safe estimates
summarized in Table~\ref{tab:summarycombinedtimes} show that for a
depth-83 circuit, this experiment could take $\approx 3.5$ days on
Vulcan, but would be completed in less than a day on Sequoia, the
larger Blue Gene/Q supercomputer at Lawrence Livermore National
Laboratory, or on Summit, the IBM-Power9/NVIDIA-Volta supercomputer
recently installed at the Oak Ridge National Laboratory.
Table~\ref{tab:summarypercentIO} further summarizes the estimated
percentage of time spent performing secondary storage access.
Remarkably, we estimate that less than half the time would be spent
performing disk I/O on the older Sequoia and Vulcan supercomputers,
and that these estimates drop to around half of that for the newer
Summit file system and to less than four percent using the Summit
solid-state burst buffer.  Importantly, we also describe in the
Supplementary Information how this simulation approach can be
generalized to handle arbitrary circuits and not merely universal
random circuits.

\begin{table}[tbp]
  \centering
  \begin{tabular}{|l|c|c|c|c|c|}
    \hline
    && Depth 55 & Depth 55 & Depth 83 & Depth 83 \\
    && Single & Double & Single & Double \\
    & Size & Precision & Precision & Precision & Precision \\
    Storage System & (PB) & (hours) & (hours) & (hours) & (hours) \\
    \hline
    Summit Burst Buffer & 7 & 7.16 & --- & 11.58 & ---\\
    Summit File System & 250 & 7.97 & 9.00 & 13.19 & 15.26 \\
    Sequoia File System & 50 & 9.68 & 12.42 & 16.60 & 22.09 \\
    Vulcan File System & 5 & 48.92 & --- & 86.85 & --- \\
    \hline
  \end{tabular}
  \caption{Estimates of total run times, including secondary storage
    read/write times of various precisions, for the computation of
    full quantum state vectors of deep $7 \times 7$-qubit circuits.}
  \label{tab:summarycombinedtimes}
\end{table}

\begin{table}[tbp]
  \centering
  \begin{tabular}{|l|c|c|c|c|}
    \hline
    & Depth 55 & Depth 55 & Depth 83 & Depth 83 \\
    & Single & Double & Single & Double \\
    & Precision & Precision & Precision & Precision \\
    Storage System & (\%) & (\%) & (\%) & (\%) \\
    \hline
    Summit Burst Buffer & 3.18 & --- & 3.93 & --- \\
    Summit File System & 12.98 & 22.98 & 15.68 & 27.12 \\
    Sequoia File System & 28.33 & 44.15 & 33.02 & 49.65 \\
    Vulcan File System & 43.30 & --- & 48.78 & --- \\
    \hline
  \end{tabular}
  \caption{Estimates of the percentage of time spent performing
    secondary storage read/write operations during the computation of
    full quantum state vectors of $7 \times 7$-qubit circuits of
    different depths, according to the desired numerical precision.}
  \label{tab:summarypercentIO}
\end{table}

\subsection{Analysis of circuit simulation difficulty}
The results reported in the previous section are based on circuits
generated according to the rules described in
\cite{boixo2018supremacy}.  Subsequently,
\cite{markov2018quantum,boixo2018repository} introduced several
changes to the benchmark circuits, making them harder to simulate
\red{using the tensor-network-based methodologies known up to that
  point in time}.  In this section we present a quantitative
assessment of the impact of these changes. \red{We remark that some of
  these changes may not affect the difficulty of simulation when using
  techniques that are not based on tensor networks, e.g.,
  Clifford-based approaches \cite{bravyi2018simulation}.}

\begin{figure}[tb]
  \centering
  \includegraphics[width=1.0\textwidth]{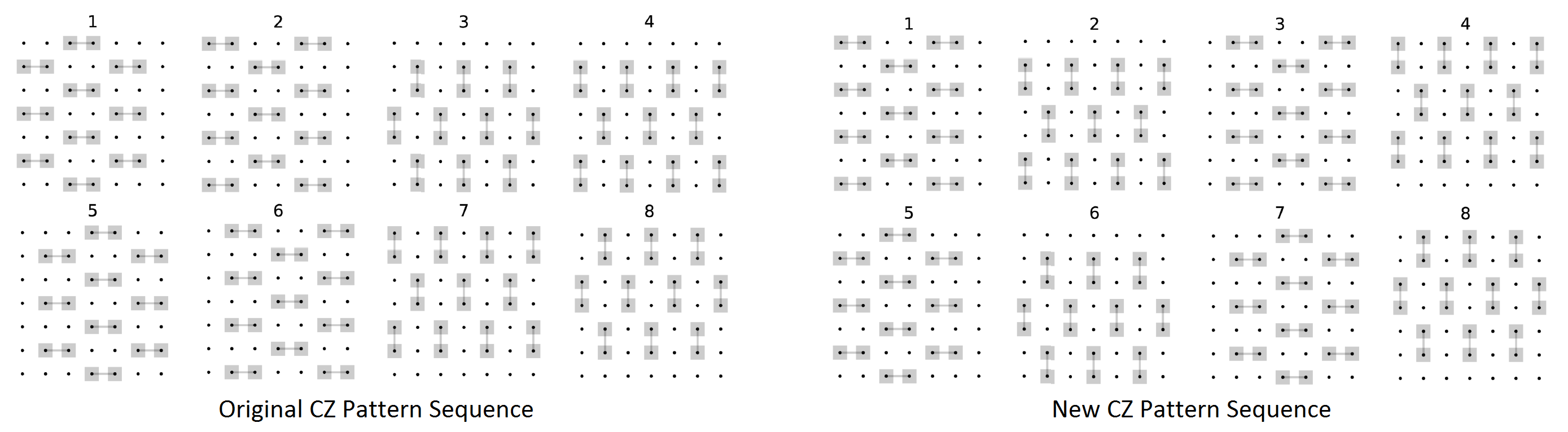}
  \caption{{\em (a) Left.} The sequence of patterned layers of
      $CZ$-gate defined by the rules described in
      \cite{boixo2018supremacy}. {\em (b) Right.} The corresponding
      pattern sequence used in the circuits that are available in the
      GRSC public repository \cite{boixo2018repository}.}
  \label{fig:newoldczpatterns}
\end{figure}

Three structural differences exist between the revised circuit
generation rules and the original rules:
\begin{itemize}
\item The addition of a final layer of Hadamard gates;
\item A different ordering of the patterned layers of $CZ$ gates, as
  illustrated in Fig.~\ref{fig:newoldczpatterns}\footnote{There
      seems to be a small discrepancy between the pattern of the $CZ$
      gates reported in \cite{villalonga2018flexible}, and the
      circuits that are available in \cite{boixo2018repository}. We
      use the pattern from \cite{boixo2018repository}.}; and
\item A different rule to randomly insert $X^{1/2}$, $Y^{1/2}$, and
  $T$ single-qubit gates, to disallow sequences of $CZ-T-CZ$ gates
  from occurring along qubit lines.
\end{itemize}
Under both the original and revised rules, every circuit begins with a
layer of Hadamard gates followed by layers of $CZ$ gates, patterned as
shown in Fig.~\ref{fig:newoldczpatterns} according to the original or
revised rules. Under the original rules, single-qubit gates are
inserted as follows:
\begin{itemize}
\item Wherever possible, insert a single-qubit gate on a qubit line
  immediately after a $CZ$ gate subject to the following constraints:
  \begin{itemize}
  \item If the most recent single-qubit gate applied to that qubit is
    a Hadamard gate, then insert a $T$ gate;
  \item Otherwise, randomly select the gate to insert from the set
    $\{X^{1/2}, Y^{1/2}, T\}$ subject to the constraint that the
    selected gate must be different from the most recent single-qubit
    gate applied to that qubit.
  \end{itemize}
\end{itemize}
Under the revised rules, single-qubit gates are inserted as follows
\cite{villalonga2018flexible}:
\begin{itemize}
\item Wherever possible insert a single-qubit gate randomly selected
  from the set $\{X^{1/2}, Y^{1/2}\}$ on a qubit line
  immediately after a $CZ$ gate.
\item After performing the above insertions, wherever possible insert
  a $T$ gate on a qubit line immediately after a $X^{1/2}$, $Y^{1/2}$,
  or $H$ gate.
\end{itemize}

Viewing each structural change as an independent experimental
treatment, we employ a full-factorial experimental design to evaluate
the effects of all $2^3 = 8$ combinations of these treatments and the
control (i.e., the original benchmark definition). Our quantitative
assessment consists in estimating the Pareto frontiers of the
computation vs.\ memory trade-off for each experimental combination,
using the $A^*$ optimization algorithm discussed earlier.

We report analysis results for both 49- and 56-qubit circuits.  In the
49-qubit case, adding final layers of Hadamard gates to our original
depth-27 circuit, and subtracting them from the corresponding
inst\_7x7\_28\_0.txt circuit in the GRSC public repository
\cite{boixo2018repository}, provides us with four out of these eight
combinations. To generate the other four we modify our circuit
generation code, which we validate against circuits in the GRSC
repository. In the 56-qubit case, the depth-23 circuit discussed in
the previous section employs a $8 \times 7$ grid of qubits, whereas the
GRSC repository contains circuits only for $7 \times 8$ grids.  To
resolve this mismatch, we use the inst\_7x8\_24\_0.txt circuit from
the GRSC repository both with and without the final layer of Hadamard
gates, and we then generate the remaining six treatment combinations
using our modified circuit generation code.

\begin{figure}[tb]
  \centering
  \includegraphics[width=0.75\textwidth]{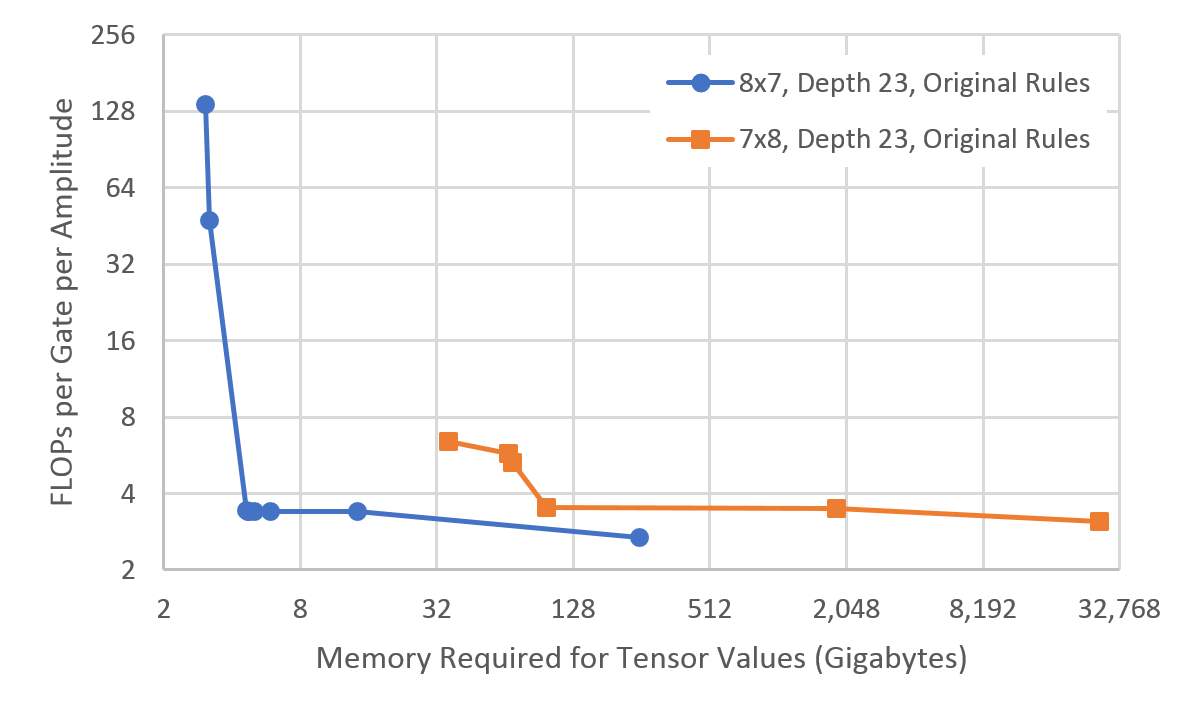}
  \caption{Pareto frontiers for $8 \times 7$- and $7 \times
      8$-qubit, depth-23 circuits generated according to the rules
      described in \cite{boixo2018supremacy}.}
  \label{fig:8x7vs7x8circuits}
\end{figure}

Somewhat surprisingly, the $7 \times 8$, depth 23 circuit generated
according to the original rules as described above is harder to
simulate than the $8 \times 7$, depth 23 circuit discussed in the
previous section, as can be seen in the Pareto frontiers shown in
Fig.~\ref{fig:8x7vs7x8circuits}. Indeed, when sufficent memory is
available the $7 \times 8$ and $8 \times 7$ circuits require roughly
the same levels of computation in terms of FLOPs per gate per
amplitude; however, as available memory is reduced, the computational
requirements of the $7 \times 8$ circuit begin to rise sharply as the
memory threshold decreases, whereas the computational requirements of
the $8 \times 7$ circuit continue to remain relatively flat. The
computational requirements for the $8 \times 7$ circuit eventually
rise sharply as well, but at a much smaller memory threshold. In the
subsequent analysis, we use the location of such bends in the Pareto
frontier to quantitatively compare the simulation difficulty of
different circuits.

\begin{figure}[tb]
  \centering
  \includegraphics[width=0.75\textwidth]{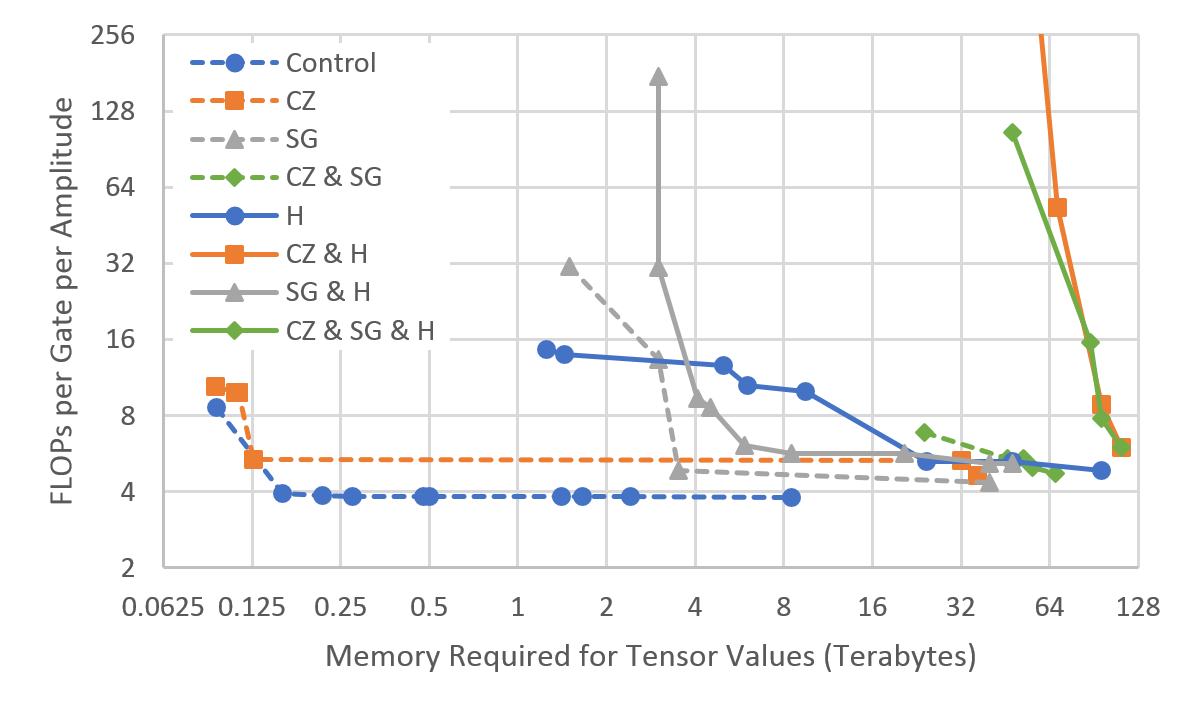}
  \caption{Pareto frontiers for the $7 \times 7$-qubit, depth-27
      circuits analyzed to assess the effects of the various
      changes in circuit generation rules introduced in
      \cite{markov2018quantum,boixo2018repository}.}
  \label{fig:7x7frontiers}
\end{figure}

\begin{figure}[tb]
  \centering
  \includegraphics[width=0.75\textwidth]{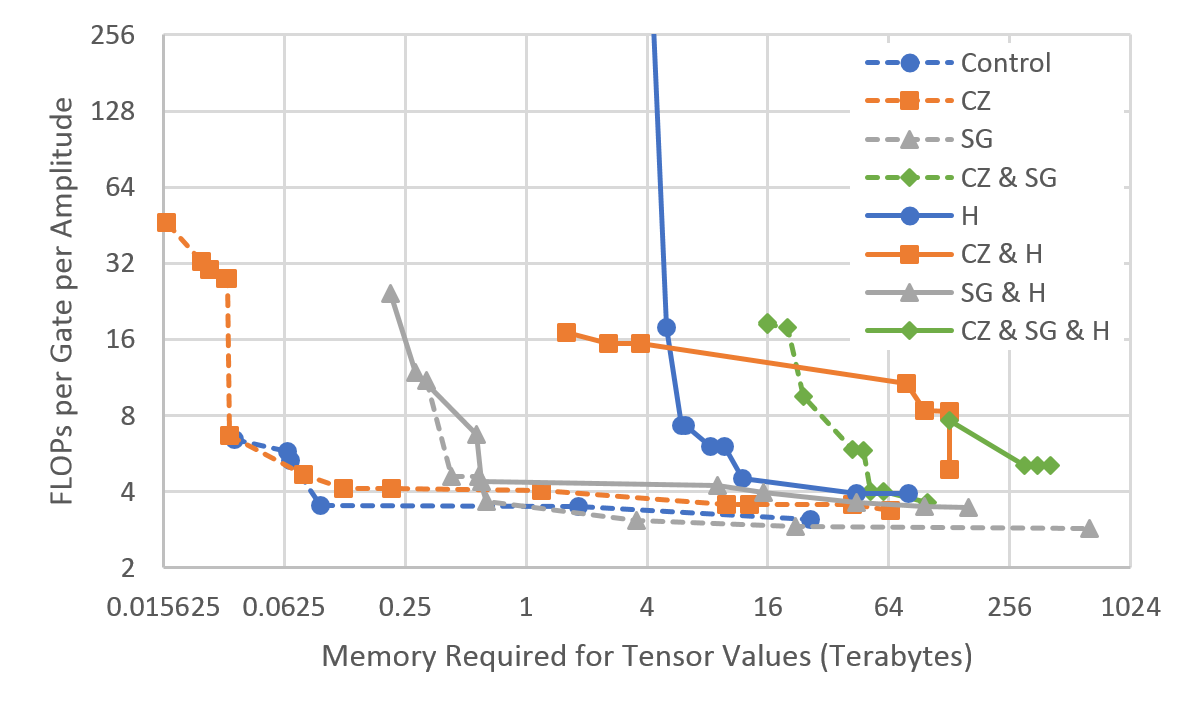}
  \caption{Pareto frontiers for the $7 \times 8$-qubit, depth-23
      circuits analyzed to assess the effects of the various
      changes in circuit generation rules introduced in
      \cite{markov2018quantum,boixo2018repository}.}
  \label{fig:7x8frontiers}
\end{figure}

The Pareto frontiers obtained for the full-factorial experimental
analysis of the $7 \times 7$ and $7 \times 8$ circuits are shown in
Figs.~\ref{fig:7x7frontiers} and~\ref{fig:7x8frontiers}, respectively.
We use the following abbreviations:
\begin{itemize}
  \item{``Control'' refers to the original circuit generation rules
    presented in \cite{boixo2018supremacy};}
  \item{``H'' refers to the experimental treatment of appending a
    final layer of Hadamard gates;}
  \item{``CZ'' refers to the experimental treatment of replacing the
    original sequence of $CZ$ layers presented in
    \cite{boixo2018supremacy} with the revised sequence used in
    \cite{boixo2018repository} as illustrated in
    Fig.~\ref{fig:newoldczpatterns}; and}
  \item{``SG'' refers to the experimental treatment of replacing the
    original rules for inserting single-qubit gates presented in
    \cite{boixo2018supremacy} with the revised rules used in
    \cite{boixo2018repository} as described above.}
\end{itemize}

We summarize the effect of the various changes in the following chain
of inequalities indicating the relative difficulty (based on the
location of the bend in the respective Pareto frontiers):
\begin{equation*}
  \text{Control}\ \lessapprox\ \text{CZ}\ <\ \text{SG}\ \lessapprox\ 
  \text{SG \& H}\ <\ \text{H}\ <\ \text{CZ \& SG}\ <\ \text{CZ \& H}\ 
  \lessapprox\ \text{CZ \& SG \& H}\ .
\end{equation*}
To conclude, all changes in the circuit generation rules tend to
increase the difficulty of simulation both individually and in
combination, but in some cases the increase is modest (e.g., the
revised sequence of $CZ$ gates), while in other cases the increase is
significant (e.g., the final layer of Hadamard gates, which appears to
be the single most impactful change).  Curiously, combining the new
single-gate rules with a final layer of Hadamard gates yields circuits
that are easier compared to inserting the Hadamards
alone. Nevertheless, the combination of all changes dramatically
increases the necessary computational resources, at least when using
the tensor network techniques discussed in this paper.  Note that all
circuits considered still lie within reach of existing supercomputers.

\section{Discussion}
This paper presents a quantum circuit simulation methodology that
introduces and extends several techniques. These techniques allowed us
to simulate universal random circuits of a size that, for some time,
were the largest to be simulated.

Among the techniques used in this paper and its arXiv preprint, a key
contribution is the concept of contraction deferral. We provide a
perspective on the recent literature in this area, highlighting the
importance of contraction deferral, especially in combination with
tensor slicing, see
\cite{chen201864,li2018quantum,chen2018classical,markov2018quantum,villalonga2018flexible,villalonga2019frontier,zhang2019classical}.
This combination has enabled the simulation of circuits much deeper
than what is presented here. Another key contribution of our paper is
the demonstration that the circuit partitioning and slicing techniques
described in \cite{haner2017simulation} can be extended to minimize
data transfer at all levels of a system memory hierarchy. With this
extension, secondary storage becomes a viable option for quantum
circuit simulation.  These techniques are not restricted to
supercomputers, but could be leveraged on conventional servers in
order to expand the range of circuits that can be routinely simulated
on those systems. An interesting question left to future exploration
is whether combining sliced and non-sliced contraction deferral both
along and across qubit lines can
provide additional benefits, and to what extent.



\subsection*{Data availability}
The datasets generated during and/or analysed during the current study
are available from the corresponding author on reasonable request.

\subsection*{Competing interests}
The authors declare that there are no competing interests.

\subsection*{Author contributions}
The first three authors contributed equally to the overall conception,
design, and drafting of the work, with E.P.\ contributing the idea of
contraction deferral, J.A.G.\ contributing refinements to the tensor
slicing methods, and G.N.\ contributing the directed hypergraph
formulation and integer programming approach.  The first four authors
contributed equally to the conception of the approach to leveraging
secondary storage.\ T.M. contributed refinements to the $A^{*}$
algorithm. E.S.\ contributed improvements to the Cyclops Tensor
Framework. E.W.D., E.T.H., and R.W.\ contributed to the planning of
the experimental setup and its deployment on Vulcan, as well as to the
analysis and interpretation of experimental results.

\section*{Acknowledgments}
We are very grateful to Jonathan L.~DuBois, Jay Gambetta, Ramis
Movassagh, John A.~Smolin, Frederick H.~Streitz, Maika Takita, Robert Walkup,
and Christopher J.~Wood. Work of Thomas Magerlein was performed during an internship
at the IBM T.J.~Watson Research Center. Work of Erik W.~Draeger and
Eric T.~Holland was performed under the auspices of the
U.S.\ Department of Energy by the Lawrence Livermore National
Laboratory under Contract No.\ DE-AC52-07NA27344. This work was
supported in part by the Laboratory Directed Research and Development
under Grant No.\ 16-SI-004.

\clearpage
\appendix
\section{Supplementary Information}
The majority of this appendix is devoted to a detailed explanation of
the mathematical foundations of our quantum circuit simulation
methodology. Some of the concepts are repeated from the main text, but
here we provide a more formal exposition that is useful to give a
precise description of the algorithms. Based on these foundations, we
describe a methodology capable of simulating deep 49-qubit circuits, up to
arbitrary depth, giving a safe estimate of the required wall-clock
time to show feasibility of such simulations. We also present an
integer programming model that describes the space of feasible
circuit partitionings considered by our methodology. The model can
be used to search for partitionings with the desired properties.
Finally, we discuss the relationship between our hypergraph
representation with that of
\cite{markov2008simulating,boixo2017simulation}.

\subsection{Preliminaries} The description of diagonal tensors in the
main text of the paper emphasizes that input index labels can carry
over to output index labels, which may lead to the same index label
appearing multiple times when there is a connected sequence of
diagonal gates. \red{For separable gates, one can use the
  corresponding diagonal gate in the tensor representation ---
  adjusting the output indices to account for the permutation whenever
  the tensor is contracted, see below.} We therefore define a {\em
  tensor network} as a hypergraph $G = (V, E, \lambda)$ consisting of:
\begin{itemize}
\item A vertex set $V$, such that each tensor is associated with a vertex.
\item A hyperedge multiset $E$, such that each hyperedge $e$ is
  an ordered subset of $2^V$.
\item A labeling function $\lambda$ that associates a unique index
  label with each hyperedge $e \in E$.
\end{itemize}
For every hyperedge $e \in E$, we call {\em tail} the first node of
$e$, {\em head} the last node of $e$.

The literature contains some examples of this type of tensor networks,
see e.g.,
\cite{buerschaper2009explicit,csahinouglu2014characterizing}, but to
the best of our knowledge these concepts were not formalized and
systematically used in the context of quantum circuit simulation.

Contractions can be defined on the type of tensor network indicated
above. Since, as is discussed in the main text, our simulation algorithm
also employs contractions between non-adjacent nodes, which are not
standard in the tensor network literature, we give our own definition
of contraction. Given a tensor network $G = (V, E, \lambda)$ and two
vertices $u,v \in V$, we define a {\em contraction} of $u,v$ as the
tensor network $G' = (V', E', \lambda')$ such that:
\begin{enumerate}[(a)]
\item $V' = V \setminus \{u,v\} \cup \{w\}$, where $w$ is a new node.
\item $E'$ is obtained from $E$ by replacing, for each hyperedge,
  each maximal occurrence of a subsequence containing only $u$ and
  $v$ with $w$. The label $\lambda(e')$ for a contracted hyperedge
  $e'$ is the same as the label $\lambda(e)$ for the original
  hyperedge $e$.
\item The tensor $T$ associated with $w$ is obtained as follows. Let
  $A$ be the tensor associated with $u$, $B$ the tensor associated
  with $v$. Let $S_{uv}$ be the set of index labels associated with
  hyperedges of the form $\{u,v\}$, $S_{vu}$ the set of index labels
  associated with hyperedges of the form $\{v,u\}$. We
    distinguish between output indices, associated with a dual space
    (e.g., the rows of a matrix), and input indices, associated with a
    primal space (e.g., the columns of a matrix). In terms of
    notation, when subscripting tensors we indicate output indices
    before the comma, and input indices after the comma.  Let $I_A$
  (resp.\ $I_B$) be the set of all input index labels for $A$
  (resp.\ $B$), $O_A$ (resp.\ $O_B$) be the set of all output index
  labels for $A$ (resp.\ $B$). Thus, the tensors $A, B$ can be written
  as:
  \[
    A_{\{k\}_{k \in S_{uv}} \{\ell\}_{\ell \in O_A
        \setminus S_{uv}}, \{i\}_{i \in S_{vu}} \{j\}_{j \in I_A \setminus
        S_{vu}}}, \qquad
    B_{\{i\}_{i \in S_{vu}} \{n\}_{n \in O_B
        \setminus S_{vu}}, \{k\}_{k \in S_{uv}} \{m\}_{m \in I_B \setminus
        S_{uv}}}.
    \]
  The tensor $T$ is then defined as:
  \begin{eqnarray}
    \label{eq:contraction}
    T_{\{\ell\}_{\ell \in O_A
        \setminus S_{uv}} \{n\}_{n \in O_B
        \setminus S_{vu}}, \{j\}_{j \in I_A \setminus S_{vu}} \{m\}_{m \in I_B \setminus
        S_{uv}}} := \\
    \sum_{\{i\}_{i \in S_{vu}}} \sum_{\{k\}_{k \in S_{uv}}} A_{\{k\}
      \{\ell\}_{\ell \in O_A \setminus S_{uv}}, \{i\}
      \{j\}_{j \in I_A \setminus S_{vu}}}
    B_{\{i\}
      \{n\}_{n \in O_B \setminus S_{vu}}, \{k\} \{m\}_{m \in I_B \setminus S_{uv}}}. \notag
  \end{eqnarray}
\end{enumerate}
We highlight some special cases of the above
definition. ``Traditional'' tensor network contractions occur when the
edges $\{u,v\}$ or $\{v,u\}$ exist. In case $S_{uv} = S_{vu} = \emptyset$,
the tensor contraction is an outer product between $A$ and $B$.
If $A$ and $B$ are diagonal, $T$ is diagonal.

It is straightforward to note that given any set of nodes, any
sequence of pairwise contractions that contracts the entire set yields
the same outcome, regardless of the order (although the computational
cost of the contraction may depend on such order). Thus, the concept
of contraction generalizes naturally to sets of vertices: given a
tensor network $G = (V, E, \lambda)$ and a subset of nodes $C
\subseteq V$, a contraction of $C$ is a tensor network $G' = (V', E',
\lambda')$ obtained by performing pairwise contractions of nodes in
$C$, in any order, until $C$ collapses to a single node.

\subsection{Simulation of quantum circuits}
A {\em quantum circuit} on $q$ qubits with depth $D$ is a tensor
network with the following properties:
\begin{itemize}
\item There are $q$ rank-1 nodes associated with the tensor
  $\ket{0}$, called the ``initial state''.
\item There are $q$ hyperedges whose head is a rank-1 node
  associated with the tensor $\bra{0}$, or a rank-1 node associated
  with the tensor $\bra{1}$, or a special node ``O'' that is not
  associated with any tensor and cannot be contracted, representing
  ``open wires.'' The heads of these $q$ hyperedges are called the
  ``output state.''
\item Every node in the graph that is not an initial or output state
  is such that its input rank is equal to the output rank.
\item All hyperedges are directed away from the initial state and
  toward the output state, i.e., there is a topological
  ordering of the graph such that the initial state precedes every
  other node, and the output state follows every other node.
\item The maximum number of nodes on any path from the initial state
  to the output state is $D + 2$.
\end{itemize}
It follows from this definition that the graph can be partitioned into
$D + 2$ sets of nodes (``layers'') with the property that each layer
has at most $q$ input hyperedges and $q$ output hyperedges (the
``qubit lines''). While we assume that every non-boundary tensor in a
quantum circuit is either a rank-2 tensor (single-qubit gate) or a
rank-4 tensor (two-qubit gate), most of our analysis directly applies
to quantum circuits with gates that involve more than two qubits.

Given a quantum circuit $G = (V, E, \lambda)$ on $q$ qubits, a {\em
  simulation strategy} for $G$ is a sequence of contractions on $G$
that contracts every node except the special node ``O.'' The node
  ``O'' is special precisely because it does not represent a tensor
  and is therefore not contracted.

We now come to defining the computational requirements for
contractions, in terms of memory occupation and number of
operations. We are interested in giving upper bounds to such
costs. For this, we have to start mixing purely theoretical
considerations with practical considerations. We assume that we have
access to both primary and secondary storage: primary storage is a
fast but more scarcely available resource, e.g., RAM, while secondary
storage is slower but available in larger capacities, e.g., disk.

Assume we are given a quantum circuit $G = (V, E, \lambda)$ and a set
of nodes $C \subseteq V$ to contract that would result in a rank $n$
tensor. Assume further that there are $m$ hyperedges that are fully
contained in $C$. The result of the contraction takes space $O(2^n)$,
and this is a space requirement that cannot be avoided. However, if
such a contraction is the final goal of the computation, then the
result of the contraction can be stored to disk rather than kept in
RAM.
Given these considerations, the {\em computational
  cost} of a simulation strategy is given by the number of floating
point operations to perform the simulation. Its {\em memory cost} is
the amount of space in primary and secondary storage that is required
to store all intermediate tensors, i.e., all tensors except the final
tensor connected to the special node ``O.''

\subsection{Asymptotic performance of quantum circuit simulation strategies}
For the sake of completeness, we discuss the asymptotic performance of
the most well-known circuit simulation strategies; the discussion in
this section considers primary storage only. There are two very
well-known ways to perform simulation of a circuit with $q$ qubits and
depth $D$, having different computational cost.

The first approach, sometimes called the ``Schr\"odinger approach,''
consists of contracting the circuit in layers. Given the topological
ordering of the circuit, we can partition the circuit into $D+2$ sets
$V_0,\dots,V_{D+1}$ such that $V_0$ contains the initial state,
$V_{D+1}$ the output state, and there is no hyperedge fully
contained in each $V_d$. We first contract $V_0$, which creates a
$O(2^q)$-size tensor, and then iteratively contract the newly created
node with $V_{d}$, one node at a time, for $d=1,\dots,D+1$. Since
there are no interior hyperedges for any of the sets $V_d$, every
contraction has memory and computational cost $O(2^q)$, for a total of
$O(D2^q)$ flops and $O(2^q)$ memory.

The second approach, sometimes called the ``Feynman approach,''
consists of contracting the entire network in one step, producing
  the output amplitudes one at a time by contracting all interior
  hyperedges. This amounts to evaluating each amplitude by summing
  over all possible paths. Assume that there are $n$ open output
wires, i.e., hyperedges whose head is ``O.'' Since the number of
hyperedges in the circuit is $O(qD)$, the total computational cost is
then $O(2^{n+qD})$, whereas memory occupation is $O(nD)$.

Recent work in the area provides better upper bounds. The paper
\cite{markov2008simulating} shows that that the simulation of quantum
circuits can be performed in time exponential in the treewidth of the
underlying tensor network representation. The setting of
\cite{markov2008simulating} assumes that there is a full set of $q$
output states $\bra{0}$ or $\bra{1}$, i.e., we are interested in
computing only one of the state amplitudes at the end of the circuit.
In this setting, simulating a circuit amounts to contracting the
entire tensor network to a single node, obtaining a scalar. To this
end, the paper \cite{markov2008simulating} proposes a simulation
strategy that always contracts adjacent nodes, according to an order
obtained via a tree decomposition of the line graph of the tensor
network. This yields a simulation strategy with cost (both
computational and memory) exponential in the treewidth of said graph,
which is within a multiplicative factor of the treewidth of the tensor
network. For a discussion of treewidth and tree decompositions in
combinatorial optimization, see \cite{bienstock1995algorithmic}.

The paper \cite{aaronson2016complexity} introduces a simulation
strategy that yields the best known upper bounds to date. For a
general circuit, by recursively splitting the circuit in half
according to the layers (i.e., creating two circuits of depth $D/2$,
then four circuits of depth $D/4$, and so on),
\cite{aaronson2016complexity} gives an algorithm that takes
$O(q(2D)^{q+1})$ time and $O(q \log D)$ memory. Specializing the
algorithm to quantum circuits with an underlying two-dimensional grid
connectivity graph, the running time decreases to $O\left(2^n \left[1 +
  \left( \frac{D}{c\sqrt{q}} \right)^{q+1} \right]\right)$ and the space
requirement increases to $O(D q \log q)$. To accomplish this goal,
\cite{aaronson2016complexity} first recursively partitions each layer
of the circuit, according to its grid structure, in smaller grids with
few edges across the sets of the partition, called a cut set. Because
of the grid structure, the number of edges in the cut set is
$O(\sqrt{q})$. Then, each set of of the partition is equivalent to a
depth $D$ circuit with a small number of qubits, and these subcircuits
can be contracted independently, provided we allow for extra ranks in
the tensors to account for the cut-set, and for the computational cost
of the final contraction of edges in the cut-set. Rather than
performing a full contraction of the subcircuits in one step,
\cite{aaronson2016complexity} recursively splits the circuit in half
according to the layers, as mentioned above. The combination of these
two recursive splitting ideas (splitting qubits according to the grid
structure, and splitting the layers in half) gives a simulation
strategy with the specified runtime.

\subsection{Slicing}
\label{s:slicing}
Assume we are given a quantum circuit $G = (V, E, \lambda)$ and a set
of nodes $C \subseteq V$ with $m$ interior hyperedges and $n$
hyperedges that contain at least one node in $C$ and one node outside
$C$. Choose $s < n$ hyperedges that leave $C$, and assume that we want
to perform a contraction of $C$ with some other set of tensors,
$C'$. Trivially, we can contract $C$ with computational cost
$O(2^{n+m})$ and total memory cost $O(2^n + m)$, create a rank $n$
tensor, and then perform the contraction with $C'$.

An alternative way to perform computation that takes advantage of
parallel computing capabilities is usually referred to as {\em
  slicing}. The idea is as follows: for all the $2^s$ possible values
of the chosen $s$ hyperedges, contract $C$ with the value for these
hyperedges fixed, creating $2^s$ ``sliced'' tensors. Calculating each
of the sliced tensors has computational cost $O(2^{n+m-s})$ and memory
(primary storage) cost $O(2^{n-s} + m)$. We can then contract each of
the sliced tensors with $C'$. We remark that there are $2^s$ sliced
tensors, and if we multiply the costs to compute the sliced tensors by
$2^s$ we obtain essentially the initial cost for the contraction of
$C$. However, there are two potential advantages to be gained through
slicing: the more obvious one is that the $2^s$ computations can be
performed in parallel; the less obvious, but equally important, is
that we may be able to reorder the tensor computation in such a way
that no more than a few slices of size $O(2^{n-s} + m)$ have to
be in primary storage at the same time --- the smaller the slices, the
higher in the memory stack they can reside. Our simulation scheme
makes this explicit by periodically storing or reading slices to/from
(primary or secondary) storage. An appropriate slicing strategy can
lead to significant memory savings by essentially splitting large
tensors into smaller parts that can be processed separately.

\subsection{Our simulation strategy}
\label{s:pednotron}
Given the framework defined above, the simulation strategy that we
propose consists of partitioning the tensor network associated with
the circuit into a small number of tensors with manageable rank and
few hyperedges between tensors. Within each of these tensors,
corresponding to subcircuits, we perform contraction in layers in a
manner similar to that of the Schr\"odinger strategy. Below we provide
a more formal description. We assume here for simplicity that
all output wires of the tensor network are open, i.e., connected to
``O,'' to output the full quantum state; if this is not the case, the
upper bounds on memory costs given in this section can potentially be
tightened depending on the structure of the quantum circuit. In terms
of notation, given $W \subseteq V$, we define
\[
  {\cal C}[W] := \{e \in E : \exists u,v \in e \textrm{ such that } u
  \in W, v \not\in W \},
\]
i.e., the set of hyperedges crossing the boundary of $W$. With a
slight abuse of notation, given $F \subseteq E$ and $W \subseteq V$,
we denote by $W \setminus F := W \setminus \bigcup_{e \in F} e$; in
other words, subtracting a set of edges $F$ from a set of nodes $W$
implies deleting from $W$ the nodes appearing in $F$.

Partition the vertex set $V$ of $G$ into $V_1,\dots,V_k$, which we
call ``subcircuits''. For $i=1,\dots,k$, let $E_i := {\cal C}[V_i]$ be
the set of hyperedges that contain at least one node in $V_i$, but are
not fully contained in $V_i$. The computation is divided into $t \ge
1$ steps, with input/output operations from/to secondary storage
between consecutive steps (although as we will discuss later, these
operations may be skipped if all the intermediate tensors fit in
primary storage). The sequence $\sigma_1,\dots,\sigma_k$ is a {\em
  computation order with $t$ steps} if it is nondecreasing,
$\bigcup_{i=1,\dots,k} \sigma_i = \{1,\dots,t\}$, and for every
hyperedge $e$, denoting by $\{V_j\}_{j \in \Sigma_e}$ the ordered
sequence of sets $V_1,\dots,V_k$ with nonempty intersection with $e$,
then the sequence $\{\sigma_j\}_{j \in \Sigma_e}$ is also
nondecreasing. The numbers $\sigma_1,\dots,\sigma_k$ are called the
{\em computation steps} of sets $V_1,\dots,V_k$. Since $V_1,\dots,V_k$
is a partition, we can unambiguously assign to each node in $V$ the
computation step from the subcircuit it belongs to. This simplifies
our exposition. According to this definition, every hyperedge is
directed from lower-index computation steps to higher-index
computation steps. Clearly there exists a computation order for any
partition $V_1,\dots,V_k$, because $t = 1$, $\sigma_1 = \dots =
\sigma_k = 1$ is valid. The special node ``O'' is assigned the final
computation step index $t + 1$.  For every $i=1,\dots,k, v \in V_i$,
we define $V(v) := V_i$ and $\sigma(v) = \sigma_i$.

The computation proceeds following the computation order
$\sigma_1,\dots,\sigma_k$. Subcircuits contracted at a given
computation step can be used to initialize tensors in subsequent
computation steps if they meet certain requirements. Formally, we
assume that a set $P \subset \{(i, j) : \sigma_j = \sigma_i + 1,
i,j=1,\dots,k\}$ of precedence relations is given:
\[
  V_i \prec V_j \quad \forall (i,j) \in P.
\]
The relation $V_i \prec V_j$ implies that subcircuit $V_i$ can be used
to initialize subcircuit $V_j$, therefore when $V_j$ is processed the
only open ranks in $V_i$ must be those of the hyperedges that connect
it to $V_j$. Let
\[
  \Delta := \left\{e \in E : \exists u,v \in e \textrm{ such that }
  \sigma(v) > \sigma(u) \textrm{ and } V(u) \not\prec V(v) \right\}.
\]
In other words, $\Delta$ is the set of hyperedges linking subcircuits
in different computation steps that do not follow the $\prec$
relation. We give below in Algorithm \ref{alg:sim} a pseudocode
description of our simulation strategy. In the description, we call
$S$ the set of hyperedges sliced up until that point. All edges in
$\Delta$ are sliced in the course of the algorithm.

\begin{algorithm}[h!]
  \caption{Outline of the simulation strategy.}
  \label{alg:sim}
  \begin{algorithmic}[1]
    \STATE $S \leftarrow \Delta \cap E_1$.
    \FOR{$i=1,\dots,k$}
    \FOR{all combinations of values for the hyperedges in $S$}
    \STATE \label{st:vic} contract $V_i$, loading the initial state vector
    from secondary storage and following the Schr\"odinger method
    \IF{$\sigma_{i+1} > \sigma_i$}
    \STATE \label{st:finalc} contract all tensors in $\{V_j :
    \sigma_j = \sigma_i\}$ and store the resulting tensor to secondary
    storage
    \ENDIF
    \ENDFOR
    \STATE $S \leftarrow S \cup (\Delta \cap E_{i+1})$
    \ENDFOR
  \end{algorithmic}
\end{algorithm}

We now analyze the computational and storage requirements of the
crucial steps of the algorithm. 

\noindent {\bf Proposition:} At step \ref{st:vic} of Algorithm
\ref{alg:sim}, let
\[
  m_i := \left|{\cal C}[V_i \setminus \Delta] \setminus \bigcup_{j :
    \sigma_j < \sigma_i} E_j\right|,
\]
and $\ell_i$ be the number of layers within $V_i$ in the topological
ordering of the circuit $G$. Then step \ref{st:vic} can be executed in
$O(\ell_i 2^{m_i})$ time requiring $O(2^{m_i})$ primary storage.

\noindent {\em Proof.}  Consider the first layer of $V_i$ in the
topological ordering of the graph. By definition, there are at most
$|{\cal C}[V_i]|$ hyperedges crossing the boundary of such
layer. Since edges in $\Delta$ are sliced (i.e., their value is fixed
in every iteration of the for loop at line 3), the tensors
appearing in $\Delta$ are one-dimensional and can be contracted in
linear time without adding open ranks. We are left with the tensors in
$V_i \setminus \Delta$, which can be contracted to a tensor of rank
${\cal C}[V_i \setminus \Delta]$. However, by construction all edges
in $\bigcup_{j : \sigma_j < \sigma_i} E_j$ (i.e., edges coming from
earlier layers of the circuit -- notice that sliced edges are removed
only once) connect $V_i$ to a tensor that has already been fully
contracted, and can be considered as the ``initial state'' of the
first layer of $V_i$. Thus, the first layer of $V_i$ is a tensor of
rank at most $m_i$, which takes at most $O(2^{m_i})$ primary storage
and $O(2^{m_i})$ time to construct. We can then proceed following the
topological ordering, contracting each layer with the subsequent
layer. Each such step requires $O(2^{m_i})$ time, and yields a tensor
of size $O(2^{m_i})$. Overall, this implies that $V_i$ can be
contracted in time $O(\ell_i 2^{m_i})$ requiring $O(2^{m_i})$ primary
storage, as indicated. \rule[0.4pt]{4pt}{4pt}

\noindent {\bf Proposition.}  At step \ref{st:finalc} of Algorithm
\ref{alg:sim}, let $C_i := \{V_j : \sigma_j = \sigma_i\}$ be the set
of tensors to be contracted, $\Gamma_i := \left\{e \in E: \exists u,v
\in e \textrm{ such that } u \in V_j, v \in V_h, j \neq h, V_j, V_h
\in C_i\right\} \setminus \Delta$ the set of hyperedges to be
contracted, and let $N_i := \left\{e \in E : \exists u,v \in e
\textrm{ such that } \sigma(u) = \sigma_i \textrm{ and } \sigma(v) =
\sigma_i + 1\right\} \setminus \Delta$ be the hyperedges acting as
input to the next computation step that are not sliced.  Then step
\ref{st:finalc} can be executed in $O(2^{|N_i|+|\Gamma_i|})$ time
requiring $O(\sum_{j\in C_i} 2^{m_j})$ primary storage and
$O(2^{|N_i|})$ secondary storage.

\noindent {\em Proof.} Consider the hyperedges leaving $C_i$. Those
hyperedges that do not intersect with subcircuits with computation
step $\sigma_i + 1$ must be in $\Delta$, hence their value is fixed by
the for loop. Thus, at every iteration of the for loop, we must
compute a tensor of rank $|N_i|$. To perform the contraction of the
tensors corresponding to the subcircuits in $|C_i|$ we typically store
the input tensors in primary storage, requiring $O(\sum_{j\in C_i}
2^{m_j})$ space, and perform $O(2^{|\Gamma_i|})$ operations per
element of the output tensor. The output can be stored to secondary
storage. This yields the memory and computation costs
indicated. \rule[0.4pt]{4pt}{4pt}

The two propositions above specify the memory and computational cost
of the two main resource-intensive steps of Alg.~\ref{alg:sim}. Notice
that the hyperedges in $\Gamma_i$ correspond precisely to the indices
whose contraction has been deferred (entanglement indices), as
discussed in the main text of the paper. Although the above
description leaves open the possibility of slicing such indices
(sliced contraction deferral), this was not done in our
experiments. In principle, it might be possible to perform the
computation in different ways that may be more efficient, hence our
analysis only states upper bounds. We remark that the stated upper
bounds hold for each iteration of the for loop for sliced hyperedges,
hence to obtain the total computational and secondary storage cost one
should multiply by $2^{|S|}$; however, primary storage requirements do
not scale exponentially with $|S|$, because at every iteration of the
for loop we can keep in memory only the tensors corresponding to the
current values for sliced edges. Furthermore, the for loop can be
parallelized very efficiently, as long as we ensure that all the
tensor slices involved in a computation are available on a given
compute node.

Based on the preceding analysis, our goal is to look for a
decomposition of $G$ into subcircuits so that the computational and
memory requirements are as small as possible. As mentioned in the
  main text of this paper, the search for such a decomposition can be
  carried out in a heuristic manner or with exact algorithms such as
  branch-and-bound; this is the topic of a subsequent section.
Typically, the bottleneck for the simulation is given by primary
storage requirements. Indeed, since each subcircuit is simulated
essentially using the Schr\"odinger method (with some additional
bookkeeping for ranks corresponding to hyperedges that connect to
subcircuits with the same computation step), primary memory is the
scarcest resource for our simulation strategy.  Finally, from a
practical point of view one should consider deviations from the
proposed strategy that better fit the available hardware.  For
example, for the simulation of the $7 \times 7$, depth-27 and $8
\times 7$, depth-23 circuits discussed in the main text we do not need
secondary storage at all: the tensors to be stored to disk all fit in
primary storage (one slice at a time), therefore we speed up the
computation by skipping the disk read/write cycle.

\subsection{An example of possible computation schemes}
\label{s:example}
We now provide a small example in order to illustrate the ideas
discussed above.  The memory and floating point requirements for the
operations that will be used in this example are given in Table
\ref{tab:memcpuops}.  Notice that these requirements assume that we
want to compute the full state vector and each complex number requires
16 bytes of memory (two double-precision floating point numbers).

\begin{table}[tb]
  \centering
  \begin{tabular}{|l|c|c|c|}
    \hline
    Gate & Multiplications & Additions & Memory [bytes] \\
    \hline
    $X$ & $2^{n+1}$ & $2^{n}$ & $2^{n+4}$ \\
    $Y$ & $2^{n+1}$ & $2^{n}$ & $2^{n+4}$ \\
    $Z$ & $2^{n}$ & $0$ & $2^{n+4}$ \\
    $H$ & $2^{n+1}$ & $2^{n}$ & $2^{n+4}$ \\
    $CX$ & $2^{n+2}$ & $3\cdot 2^{n}$ & $2^{n+4}$ \\
    $CZ$ & $2^{n}$ & $0$ & $2^{n+4}$ \\
    \hline
  \end{tabular}
  \caption{Memory requirements and number of floating point of operations for some quantum gates, when applied to a $n$-qubit quantum state ($2^n$-dimensional double precision complex vector).}
  \label{tab:memcpuops}
\end{table}

\begin{figure}[tb]
  \leavevmode
  \centering
  \Qcircuit @C=1em @R=.7em {
    \lstick{\ket{0}} & \ustick{i_0} \qw & \gate{H} & \ustick{j_0} \qw & \qw      & \qw       & \ctrl{1} & \qw      & \gate{T} &              \qw & \qw      & \qw      & \qw                 & \ctrl{1} & \ustick{j_0} \qw & \qw\\
    \lstick{\ket{0}} & \ustick{i_1} \qw & \gate{H} & \ustick{j_1} \qw & \ctrl{1} & \gate{T}  & \gate{Z} & \qw      & \gate{Y} & \ustick{k_1} \qw & \ctrl{1} & \gate{X} & \ustick{\ell_1} \qw & \gate{Z} & \ustick{\ell_1} \qw & \qw \inputgrouph{1}{2}{.8em}{\ket{\phi}}{4.8em} \inputgrouph{2}{3}{.8em}{\ket{\psi}}{-31em} \\
    \lstick{\ket{0}} & \ustick{i_2} \qw & \gate{H} & \ustick{j_2} \qw & \gate{Z} & \gate{T}  & \qw      & \ctrl{1} & \gate{X} & \ustick{k_2} \qw & \gate{Z} & \gate{T} & \qw                 & \qw      & \ustick{k_2} \qw & \qw \\
    \lstick{\ket{0}} & \ustick{i_3} \qw & \gate{H} & \ustick{j_3} \qw & \qw      & \qw       & \qw      & \gate{Z} & \gate{T} &              \qw & \qw      & \qw      & \qw                 & \qw      & \ustick{j_3} \qw & \qw \inputgrouph{3}{4}{.8em}{\ket{\xi}}{4.8em} 
  \gategroup{1}{3}{2}{9}{.6em}{--} \gategroup{3}{3}{4}{9}{.6em}{--}
  \gategroup{2}{5}{3}{5}{.6em}{.} \gategroup{1}{11}{3}{14}{1.7em}{--}
  }
  \caption{Example of a random $4 \times 1$-qubit, depth 15 circuit
    generated according to \cite{boixo2018supremacy}. The circuit can
    be contracted to 9 layers. The partitioning scheme is indicated in
    the figure.  The tensor corresponding to the entangling gate in
    the second layer is assigned to the bottom subcircuit.}
  \label{fig:googleex}
\end{figure}

We consider a universal random circuit for a $4 \times 1$ grid of
qubits with depth 15. It turns out that the generating rules produce
many empty layers for this example and the circuit can be represented
using 9 layers.  Such a circuit is depicted in
Fig.~\ref{fig:googleex}, together with a possible partitioning scheme.
Using the notation described in the previous section, we have three
subcircuits $V_1,V_2,V_3$ labeled $\ket{\phi}, \ket{\xi}, \ket{\psi}$
in Fig.~\ref{fig:googleex}. The computation order is $\sigma_1 = 1,
\sigma_2 = 1, \sigma_3 = 2$; the precedence relations are $V_1 \prec
V_3, V_2 \prec V_3$. The tensor for the first $CZ$ gate in the circuit
is assigned to $V_2$, and there are three sliced hyperedges (indices
$j_0, k_2, j_3$). To give a clearer description of how the proposed
algorithm works, we give the full equations governing the state using
the labeling indicated in the diagram (for simplicity, we use gate
names to indicate the corresponding matrices):
\[
  \begin{array}{rl}
  \phi_{j_0k_1j_1} &= T_{j_0,j_0} Y_{k_1,j_1} CZ_{j_0j_1,j_0j_1} T_{j_1,j_1} \sum_{i_0 \in \{0,1\}} H_{j_0,i_0} \delta_{i_0} \sum_{i_1 \in \{0,1\}} H_{j_1,i_1} \delta_{i_1} \\
  \xi_{j_1 k_2 j_3} &= T_{j_3,j_3} \sum_{j_2 \in \{0,1\}} X_{k_2,j_2} CZ_{j_2j_3,j_2j_3} T_{j_2,j_2} CZ_{j_1j_2, j_1j_2} \sum_{i_2 \in \{0,1\}} H_{j_2,i_2} \delta_{i_2} \sum_{i_3 \in \{0,1\}} H_{j_3,i_3} \delta_{i_3} \\
  \psi_{j_0 \ell_1 k_2 j_3} &= CZ_{j_0\ell_1, j_0\ell_1} T_{k_2,k_2} \sum_{k_1 \in \{0,1\}} X_{\ell_1,k_1} CZ_{k_1k_2, k_1k_2} \sum_{j_1 \in \{0,1\}} \phi_{j_0 k_1 j_1} \xi_{j_1 k_2 j_3}.
  \end{array}
\]
Note that, although the first $CZ$ gate in the circuit is assigned to
the bottom $\ket{\xi}$ subcircuit, we have to keep track of the extra
$j_1$ index associated with that $CZ$ gate in the top $\ket{\phi}$
subcircuit once the $Y$ gate is introduced.  In other words, $j_1$
becomes an entanglement index that cannot be contracted away until the
tensors corresponding to the top $\ket{\phi}$ and bottom $\ket{\xi}$
subcircuits are combined: the precedence relations $V_1 \prec V_3, V_2
\prec V_3$ ensure that this is carried out before $V_3$ is computed.
Because the hyperedges corresponding to indices $j_0, k_2, j_3$ are
sliced, we never have to explicitly materialize the full state
$\ket{\psi}$ in memory, and can work on slices. In this example, the
final state can be computed in $2^3 = 8$ slices using a state vector
of size $2^1$ for the computation of a slice, never allocating memory
to hold a full $2^4$-dimensional state vector. Furthermore, we can
decide to keep each slice in primary storage skipping all the
secondary storage input/output operations indicated in
Alg.~\ref{alg:sim}, as the size of an individual slice does not
exceed the available primary storage space.

\begin{sidewaysfigure}
  \centering
  \includegraphics[width=0.49\textwidth]{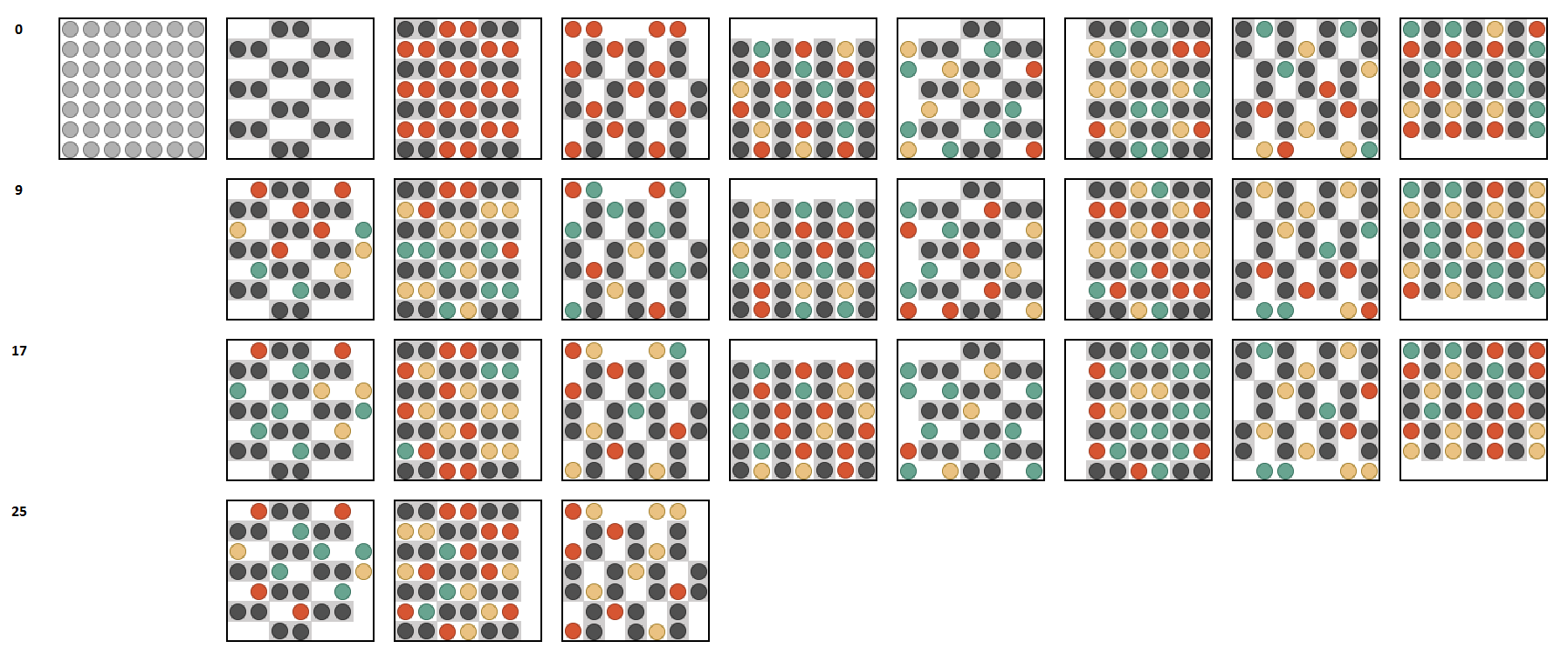}
  \includegraphics[width=0.49\textwidth]{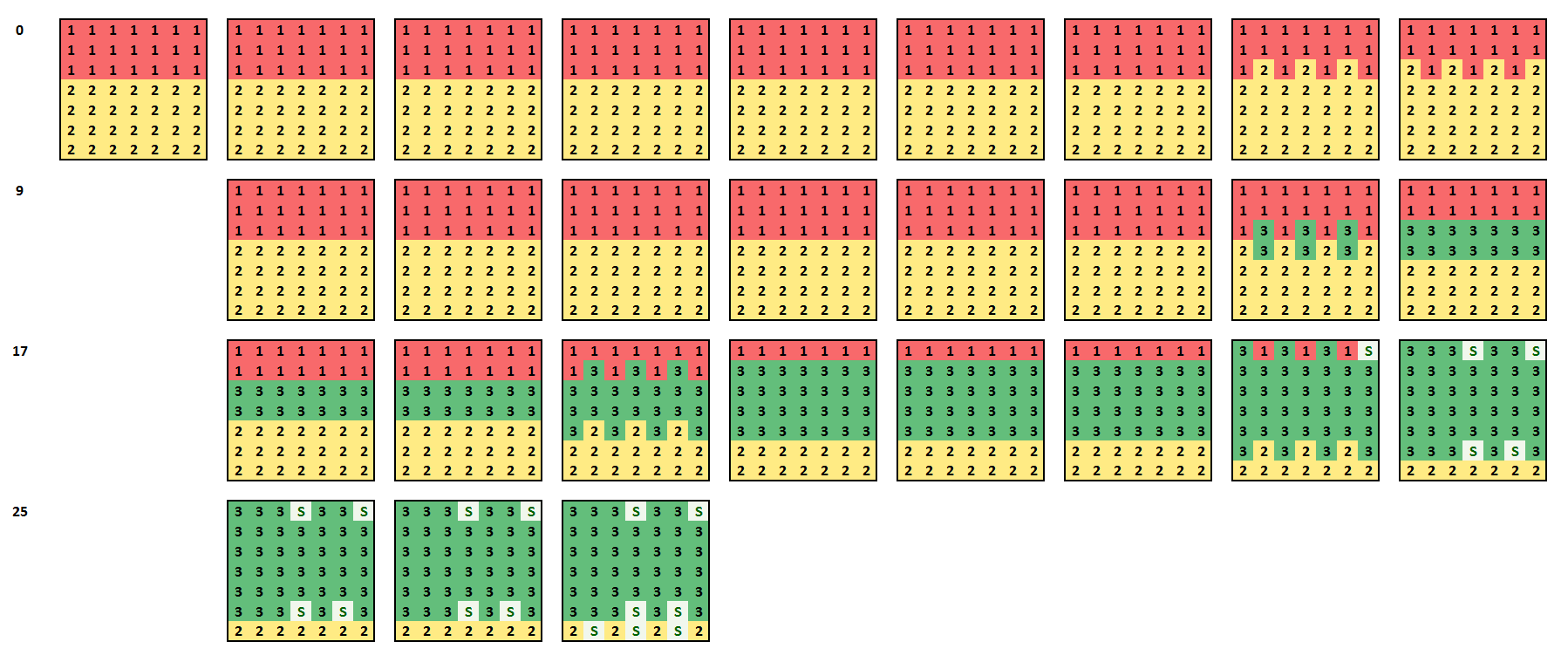}\\[5em]
  \includegraphics[width=0.49\textwidth]{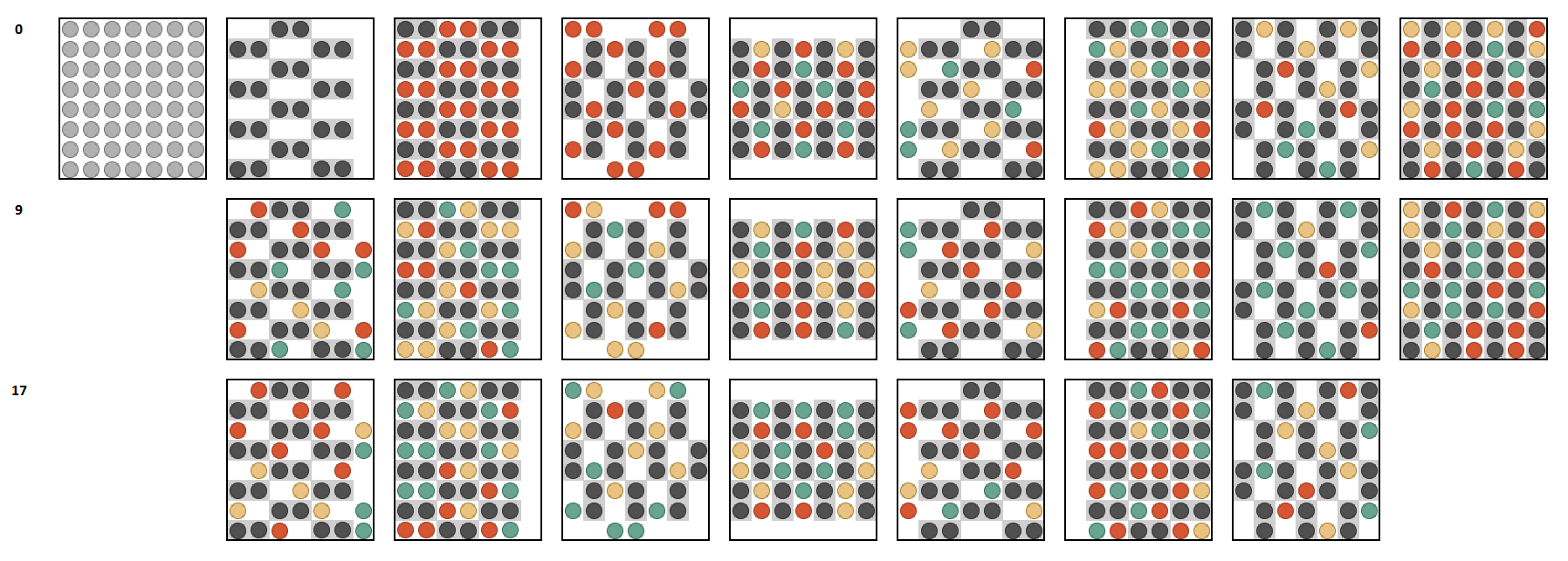}
  \includegraphics[width=0.49\textwidth]{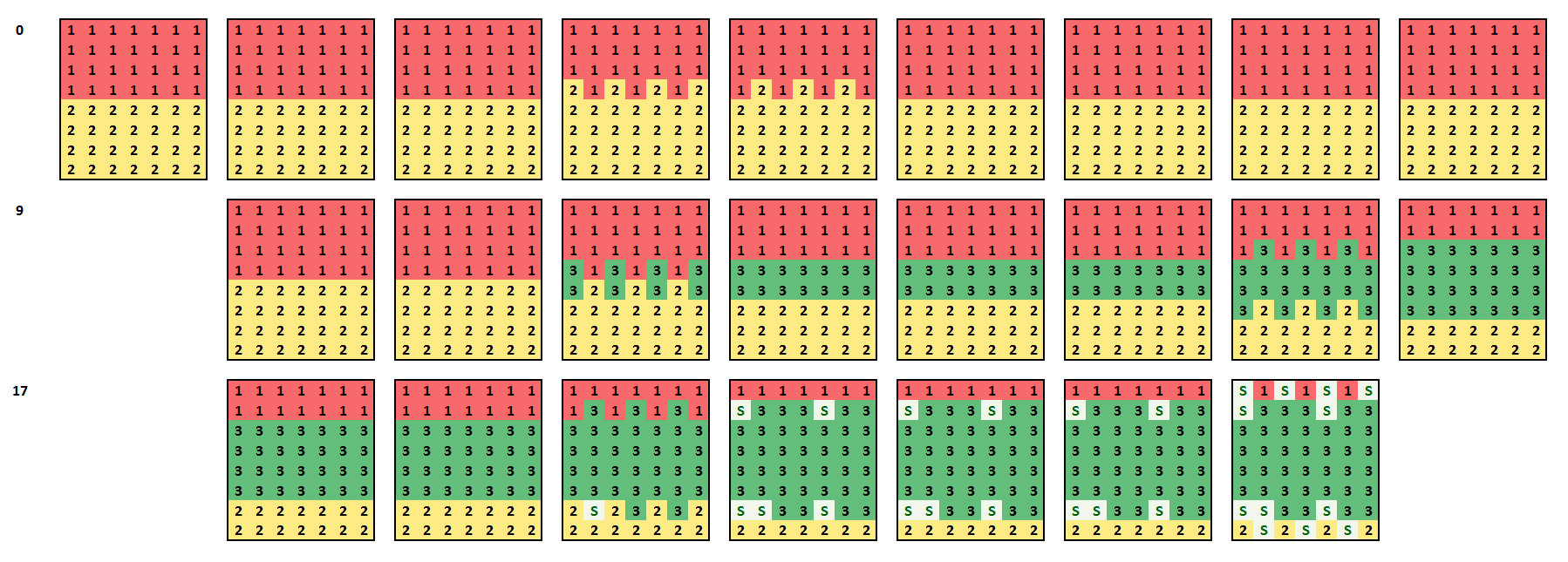}
  \caption{The 49-qubit, depth 27 circuit and the 56-qubit, depth 23
    circuit are represented on the left. The figure on the right
    contains the respective partitioning. Qubits marked with an ``S''
    can be sliced, as can the qubits in subcircuits 1 and 2 that are
    not involved in subcircuit 3 (i.e., those for which subcircuit 3
    does not contain any gates applied to them).}
  \label{fig:googlepart}
\end{sidewaysfigure}

The partitioning scheme for the $7 \times 7$, depth-27 and $8 \times 7$,
depth-23 circuits is akin to the above example. With a similar graphic
scheme (some details are omitted due to the circuit size), we depict
in Fig.~\ref{fig:googlepart} the partitioning scheme used for the
depth-27 and depth-23 simulations discussed in the main text of the
paper. We remark that these circuits are constructed according to
  the rules of the first version of \cite{boixo2018supremacy}. A
  revised benchmark was subsequently proposed \cite{boixoaiblog},
  including circuits that are more difficult to simulate; at that
  time, our manuscript was already posted online and submitted for
  publication, therefore we did not run experiments on the revised
  benchmark.

\subsection{Determining a computation scheme via integer programming}
The problem of identifying a suitable computation scheme can be
formulated as a discrete optimization problem. This is also the case
for other simulation approaches discussed in the literature; most
notably, papers based on tensor network contractions
\cite{markov2008simulating} require a treewidth decomposition of the
circuit graph. In practice, such a decomposition is computed using
heuristic or branch-and-bound algorithms; e.g.,
\cite{boixo2017simulation} reports running the branch-and-bound
algorithm QuickBB ``for a day'' (quoted) to find a suitable
decomposition. Here, we advocate an approach based on integer
programming \cite{nemhauserwolsey}. The main benefit of doing so is
that there exist off-the-shelf software packages to solve discrete
optimization problems articulated as integer programming
problems. Thus, once the mathematical model for the problem is
defined, a computation scheme can be determined automatically using
such software. We now describe the integer programming model, and
report a short summary of our computational experience.

Using notation from previous sections, we are given a hypergraph $G =
(V, E)$ describing a circuit and our goal is to find (up to) $k$
subcircuits $V_1,\dots,V_k$. For any positive integer $n$, we denote
by $[n]$ the set $\{1,\dots,n\}$. The sequence
$\sigma_1,\dots,\sigma_k$ of $t$ computation steps is assumed to be
given; this is without loss of generality because we can always pick
$k$ large and let some of the subcircuits be empty. We denote $T_c :=
\{j \in \{1,\dots,k\} : \sigma_j = c\}, c \in [t]$. For simplicity, we
assume here that the set of precedence relations is fixed and equal to
$P := \{(i,j) : \forall c \in [t-1], \forall i \in T_c, \forall j \in
T_{c+1}\}$, which is what we use in all the examples and numerical
tests discussed in this paper. It is not difficult to amend the
integer programming model described below in a way such that the
precedence relations are determined by an additional set of decision
variables.

The decision variables of the integer programming model are:
\begin{itemize}
\item $x_{i j} \in \{0,1\} \; \forall i \in [n], j \in [k]$. We write
  $x_{i j} = 1$  if $i \in V_j$, 0 if not.
\item $s_e \in \{0,1\} \; \forall e \in E$. We write $s_e = 1$ if edge
  $e \in E$ is sliced, 0 if not.
\item $s_{ec} \in \{0,1\} \; \forall e \in E, c \in [t-2]$. We write
  $s_{ec} = 1$ if edge $e \in E$ connects computation step $c$ to
  computation step $c + 2$ or greater, for $c \in [t-2]$, 0 if
  not. This is an auxiliary variable used to model $s_e$.
\item $r_{ej} \in \{0,1\} \; \forall e \in E, j \in [k]$. We write
  $r_{ej} = 1$ if edge $e \in E$ connects subcircuit $j$ to a
  different subcircuit and must be included in rank computation, 0
  otherwise.
\item $m_j \in \mathbb{Z} \; \forall j \in [k]$. The variable $m_j$
  indicates the rank of subcircuit $V_j$.
\item $z_c \in \mathbb{Z} \; \forall c \in [t]$. The variable $z_c$
  denotes the (ceiling of the) $\log$ of the complexity of computation
  step $c$ (see step \ref{st:finalc} of Algorithm \ref{alg:sim}).
\end{itemize}
To describe the space of possible circuit partitionings, we use the
following set of constraints:
\begin{itemize}
\item $\forall i \in [n]: \sum_{j=1}^k x_{i j} = 1$. Each tensor (node in the graph) belongs to exactly one subcircuit.
\item $\forall e \in E, \forall c \in [t-2]: 2s_{ec} \le \sum_{j \in
  T_c} x_{\text{tail}(e) j} + \sum_{j \in \bigcup_{h \in
    [c+2,\dots,t]} T_{h}} x_{\text{head}(e) j}$. Variable $s_{ec}$ can
  be 1 only if hyperedge $e$ starts from tensor with computation step
  $c$ and ends at a tensor with computation step $\ge c+2$.
\item $\forall e \in E: s_e = \sum_{c \in [t-2]} s_{ec}$. Hyperedge
  $e$ is sliced if and only if it connects computation step $c$ and
  $c+2$, for some $c \in [t-2]$.
\item $\forall e = \{v_1,\dots,v_\ell\} \in E, \forall p \in [\ell-1],
  \forall c \in [t-1]: \sum_{j \in T_{c+1}} x_{v_p j} + \sum_{j \in
    \bigcup_{h \in [c]} T_{h}} x_{v_{p+1} j} \le 1$. The sequence of
  computation steps along any hyperedge must be nondecreasing.
\item $\forall e = \{v_1,\dots,v_\ell\} \in E, \forall p \in [\ell],
  \forall q \in [\ell] \setminus \{p\}, \forall j \in [k], \forall h
  \in [k] \setminus \{j\}: x_{v_p j} + x_{v_q h} - s_e \le r_{ej}$. If
  hyperedge $e$ contains subcircuit $j$ and $h$, and it is not sliced,
  it contributes to the rank of $V_j$. (These constraints are
  redundant when $q < p$ and $\sigma_h \ge \sigma_j$, so they can be
  dropped in that instance.)
\item $\forall e = \{v_1,\dots,v_\ell\} \in E, \forall j \in [k]:
  x_{v_1 j} + s_e \le r_{ej}$. If $\text{tail}(e) \in V_j$ and
  hyperedge $e$ is sliced, then it contributes to the rank of $V_j$
  (as there is an edge crossing the partition that would not be
  counted otherwise).
\item $\forall j \in [k]: m_j \ge \sum_{e \in E} r_{ej}$. Rank of
  subcircuit $j$.
\item $\forall c \in [t]: z_c \ge \sum_{e \in E} \sum_{j \in T_c}
  r_{ej}$. Logarithm of the complexity of computation step $c$ (see
  step \ref{st:finalc} of Algorithm \ref{alg:sim}).
\end{itemize}
We remark that many of the constraints above are expressed as
inequalities that only enforce lower bounds on the decision variables
that model the rank of the subcircuits, i.e., tensors. As a
consequence, there exist feasible solutions in which the decision
variables indicate tensor ranks that are larger than the true
value. This issue can be taken care of by minimizing all these
variables via the objective function, or by imposing upper
bounds. Notice that the model above allows many symmetric solutions
(e.g., the indices of all subcircuits belonging to the same
computation step are interchangeable). We break some symmetries by
imposing a nonincreasing ordering in the size of the subcircuits at
each computation step.

As the objective function of the problem our intent was to use $\min \max_{j
  \in [k]} m_j$, which can easily be modeled as a linear objective
function with an additional variable, writing $\min w$ subject to $w
\ge m_j \; \forall j \in [k]$. This corresponds to minimizing the rank
of the largest subcircuit. To ensure that the decision variables
modeling the tensor ranks are at their lower bound at the optimum, we
add all these variables to the objective function, but with a
smaller coefficient than the desired objective function
$\min \max_{j \in [k]} m_j$. Thus, in our numerical experiments we use
the objective function:
\begin{equation}
  \label{eq:objip}
  \min 100 (\max_{j \in [k]} m_j) + \sum_{j \in [k]} m_j + \sum_{c \in
    [t]} z_c + \sum_{j \in [k]} (\max_{i \in V} x_{ij}),
\end{equation}
where the purpose of the last term is to break ties between solutions
in favor of solutions that use fewer subcircuits. One of the main
advantage of using an integer programming model is its
flexibility, therefore many alternative objective functions are
possible; minimizing the maximum rank has the advantage that we do not
have to explicitly model the exponential growth of memory
requirements, which would be the case if we tried to minimize the
total memory consumption $\sum_{j \in [k]} 2^{m_j}$. 

We solve the integer programming model using IBM ILOG CPLEX 12.8 on a
virtual machine instantiated in the cloud. The virtual machine has 32
Intel Xeon E5-2683 v4 @ 2.10GHz cores and 64 GB RAM. In each run we impose
a limit of 5 hours. The model is given to the solver as described
above, without any further effort to generate additional valid
inequalities that could improve performance, besides those that are
automatically generated by CPLEX. We impose the additional constraint
that at most $15$ edges can be sliced, because otherwise we may obtain
solutions that have too many such edges. This can be explained in
light of the fact that sliced edges can reduce the rank of
subcircuits, and the objective function used in these experiments
considers tensor rank only, without taking the computation time into
account. Results are summarized in Table~\ref{tab:ipres}. Notice that
we report lower and upper bounds in terms of the (easier to
understand) objective function $\min \max_{j \in [k]} m_j$, i.e., the
rank of the largest tensor, rather than the modified objective
function of Eq.~\ref{eq:objip}. While the gap between upper and lower
bounds is rather large when 6 subcircuits are considered, it is
significantly smaller with 4 subcircuits, and shows that the
partitionings used in the simulations discussed in this paper are
nearly optimal. More importantly, the results confirm that the integer
programming model is a viable approach to automatically determine
computation schemes for circuits of these or similar sizes.

\begin{table}
  \centering
  \begin{tabular}{|l|c|c|c|}
    \hline
    Circuit & $k$ & Upper bound & Lower bound \\
    \hline
    49 qubit, depth 27 & 4 & 38 & 27 \\
    49 qubit, depth 27 & 6 & 27 & 9 \\
    56 qubit, depth 23 & 4 & 38 & 29 \\
    56 qubit, depth 23 & 6 & 27 & 11 \\
    \hline
  \end{tabular}  
  \caption{Summary of the results of the automatic circuit
    partitioning based on integer programming. For each circuit and a
    given maximum number of subcircuits $k$, we report the best
    solution found and the best known lower bound after 5 hours.}
  \label{tab:ipres}
\end{table}

\subsection{Computation of single amplitudes}
The discussion so far has focused on the computation of the entire
state vector, which is a natural outcome considering that our
methodology is based on applying the Schr\"odinger approach to
subcircuits. Computing a single amplitude $\bra{x}Q\ket{y}$ for given
basis states $x,y$ is, in general, a simpler task. We can exploit the
machinery described above to compute single amplitudes for circuits
that are otherwise intractable with existing methodologies. In broad
terms, we borrow inspiration from the recursive circuit partitioning
schemes proposed in \cite{aaronson2016complexity}, but we depart from
a purely recursive approach by constructing tensors for the resulting
subcircuits using the computation scheme previously discussed.

Following \cite{aaronson2016complexity}, we first partition circuits
depth-wise.  Suppose we are given a quantum circuit of depth $d$
implementing a unitary matrix $Q$. Calling $L$ the unitary
representing the first $\approx d/2$ layers of the circuit, and $R$
the unitary representing the remaining layers, we have $Q = RL$. Our
goal is to compute $\bra{x}Q\ket{y} = \bra{x}RL\ket{y}$. This can be
accomplished by computing $L\ket{y}$ and $R^\dag \ket{x}$ separately,
then combining $\left(R^\dag \ket{x}\right)^{\dag}$ and $L \ket{y}$ to
obtain $\bra{x}RL\ket{y}$. Intuitively, this corresponds to simulating
the circuit in input-to-output order for approximately half the total
depth, and in output-to-input order for the remaining layers. The
resulting tensors can then be contracted to obtain the desired
amplitude. This idea was also developed, independently, in
\cite{li2018quantum}, but we contract the resulting tensors with a
more efficient order than in \cite{li2018quantum}.

We are interested in working with circuits whose state vectors cannot
be fully stored in primary memory, so, in general, tensors
corresponding to $L\ket{y}$ and $R^\dag \ket{x}$ cannot be constructed
directly. Instead, we apply the second partitioning scheme proposed in
\cite{aaronson2016complexity}, which is to split the circuit row-wise.
Assuming an $r \times c$ grid of qubits, the subcircuits corresponding
to $L$ and $R$ would each be further partitioned into the top $\approx
r/2$ rows of gates and the bottom remaining rows of gates.  Following
the approach described earlier, entanglement gates that bridge these
top and bottom partitions would be assigned to one partition or the
other, with entanglement indices introduced as appropriate. This
assignment affects the memory requirements, therefore it is preferable
to assign bridging gates to partitions so as to minimize the resulting
memory occupation.

If $\phi$ and $\xi$ are the top and bottom tensors constructed for the
subcircuit corresponding to $L$, and if $\psi$ and $\chi$ are the
corresponding tensors for $R$, then contracting $\phi$ and $\xi$
yields $L\ket{y}$ and contracting $\psi$ and $\chi$ yields $R^\dag
\ket{x}$.  Contracting all four tensors thus yields the desired
amplitude $\bra{x}RL\ket{y}$.  Because we have complete freedom to
choose the order of contraction, we can also consider contracting
$\phi$ and $\psi$ separately from $\xi$ and $\chi$, and then
contracting the resulting tensors.  This last order of contraction has
desirable properties in terms of both memory requirements and
floating-point operations.

\begin{figure}[tb]
  \leavevmode
  \centering
  \Qcircuit @C=1em @R=.7em {
    \lstick{\ket{0}} & \ustick{j_0} \qw &\gate{H} & \ustick{i_0} \qw      & \qw & \qw       & \ctrl{1} & \ustick{i_0} \qw & \gate{T} & \ustick{i_0} \qw      & \qw & \qw & \qw      & \qw      & \ctrl{1} & \ustick{i_0} \qw & \qw & {\; \quad \bra{a_0}} \\
    \lstick{\ket{0}} & \ustick{\ell_1} \qw & \gate{H} & \ustick{k_1} \qw & \ctrl{1} & \gate{T}  & \gate{Z} & \ustick{k_1} \qw & \gate{Y} & \ustick{j_1} \qw      & \qw & \ctrl{1} & \ustick{j_1} \qw & \gate{X} & \gate{Z} & \ustick{i_1} \qw & \qw & {\; \quad \bra{a_1}} \inputgrouph{1}{2}{.8em}{\ket{\phi}}{4.8em} \inputgrouph{1}{2}{.8em}{\bra{\psi}}{-34em} \\
    \lstick{\ket{0}} & \ustick{k_2} \qw & \gate{H} & \ustick{j_2} \qw & \gate{Z} & \gate{T}  &  \qw      & \ustick{j_2} \qw & \ctrl{1} & \ustick{j_2} \qw & \gate{X} & \gate{Z} & \ustick{i_2} \qw & \gate{T} & \qw      & \ustick{i_2} \qw & \qw & {\quad \; \bra{a_2}} \\
    \lstick{\ket{0}} & \ustick{j_3} \qw & \gate{H} & \qw      & \qw & \qw       & \qw      & \ustick{i_3} \qw & \gate{Z} & \ustick{i_3} \qw & \gate{T} &  \qw      & \qw & \qw      & \qw      & \ustick{i_3} \qw & \qw & {\quad \; \bra{a_3}} \inputgrouph{3}{4}{.8em}{\ket{\xi}}{4.8em} \inputgrouph{3}{4}{.8em}{\bra{\chi}}{-34em}
  \gategroup{1}{3}{2}{7}{.6em}{--} \gategroup{3}{3}{4}{6}{.6em}{--}
  \gategroup{2}{5}{3}{5}{.6em}{.} \gategroup{1}{9}{2}{15}{.6em}{--}
  \gategroup{3}{9}{4}{14}{.6em}{--} \gategroup{2}{12}{3}{12}{.6em}{.}
  }
  \caption{Example of a partitioning scheme to compute single
    amplitudes for the circuit in Fig.~\ref{fig:googleex}, simulating
    the circuit in input-to-output and output-to-input order. The
    tensors corresponding to the two dotted entangling gates are
    assigned to the top subcircuits.}
  \label{fig:googleexsplit}
\end{figure}

Fig.~\ref{fig:googleexsplit} illustrates these ideas on the same
circuit used for one of our previous examples.  We give below the
equations describing each of the subcircuits involved in the figure.
In these equations, indices $j_2$ and $i_2$ correspond to the
hyperedges of the $CZ$ gates connecting the two wires in the middle that
remain ``open'' in the corresponding tensor (i.e., not contracted
within the tensor).
\[
\begin{array}{rl}
  \phi_{i_0k_1j_2} &= CZ_{i_0k_1,i_0k_1}T_{k_1,k_1}CZ_{k_1j_2,k_1j_2} \sum_{j_0 \in \{0,1\}} H_{i_0,j_0} \delta_{j_0} \sum_{\ell_1 \in \{0,1\}} H_{k_1,\ell_1} \delta_{\ell1} \\
  \xi_{j_2i_3} &= T_{j_2,j_2} \sum_{k_2 \in \{0,1\}} H_{j_2,k_2} \delta_{k_2} \sum_{j_3 \in \{0,1\}} H_{i_3,j_3} \delta_{j_3} \\
  \psi_{i_0k_1i_2} &= \left(\sum_{j_1 \in \{0,1\}} X_{i_1,j_1} CZ_{j_1i_2,j_1i_2} Y_{j_1,k_1} \right) T_{i_0,i_0} CZ_{i_0i_1,i_0i_1} \delta_{i_0-a_0} \delta_{i_1 - a_1} \\
  \chi_{j_2i_3i_2} &= CZ_{j_2i_3,j_2i_3} X_{i_2,j_2} T_{i_3,i_3} T_{i_2,i_2} \delta_{i_2 - a_2} \delta_{i_3 - a_3}
\end{array}
\]
Given these four subcircuits, there are two natural ways to combine
them to obtain the desired amplitude $\alpha$: we can combine the
tensors ``vertically'' first ($\ket{\phi}$ with $\ket{\xi}$, and
$\bra{\psi}$ with $\bra{\chi}$) to obtain
\[
\begin{array}{rl}
  \lambda_{i_0k_1j_2i_3} &= \sum_{j_2 \in \{0,1\}} \phi_{i_0k_1j_2} \xi_{j_2i_3} \\
  \rho_{i_0k_1j_2i_3} &= \sum_{i_2 \in \{0,1\}} \psi_{i_0k_1i_2} \chi_{j_2i_3i_2} \\
  \alpha &= \sum_{i_0,k_1,j_2,i_3 \in \{0,1\}} \lambda_{i_0k_1j_2i_3} \rho_{i_0k_1j_2i_3},
\end{array}
\]
or ``horizontally'' first ($\ket{\phi}$ with $\bra{\psi}$, and
$\ket{\xi}$ with $\bra{\chi}$) to obtain
\[
\begin{array}{rl}
  \sigma_{i_2j_2} &= \sum_{i_0,k_1 \in \{0,1\}} \phi_{i_0k_1j_2} \psi_{i_0k_1i_2} \\
  \omega_{i_2j_2} &= \sum_{j_2,i_3 \in \{0,1\}} \xi_{j_2i_3} \chi_{j_2i_3i_2} \\
  \alpha &= \sum_{i_2,j_2 \in \{0,1\}} \sigma_{i_2j_2} \omega_{i_2j_2}.
\end{array}
\]
It is easy to verify that the second approach is more efficient,
creating intermediate tensors with smaller memory footprints and
requiring fewer floating point operations.

\begin{figure}[tb]
  \centering
  \includegraphics[width=0.47\textwidth]{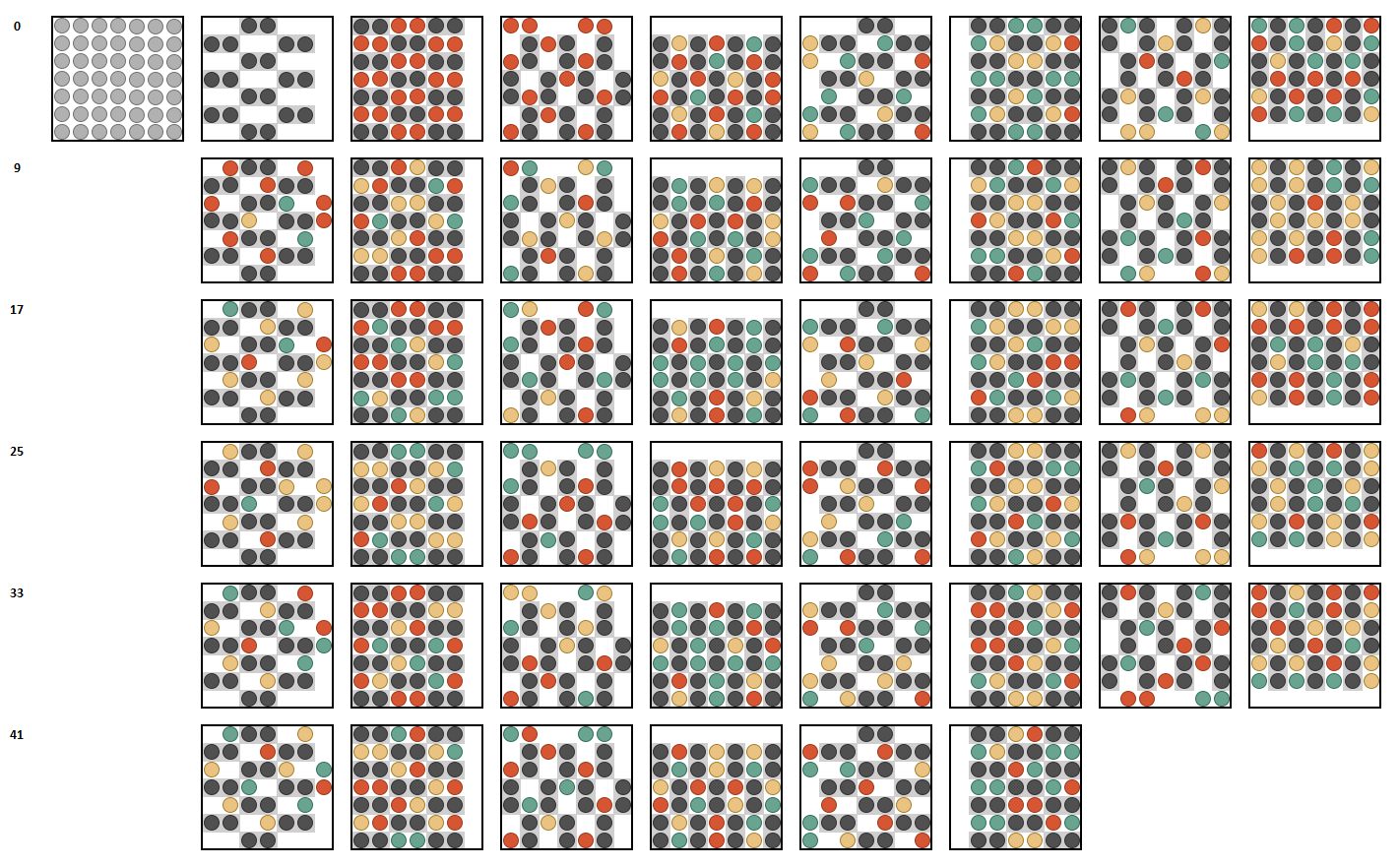}\hfill
  \includegraphics[width=0.47\textwidth]{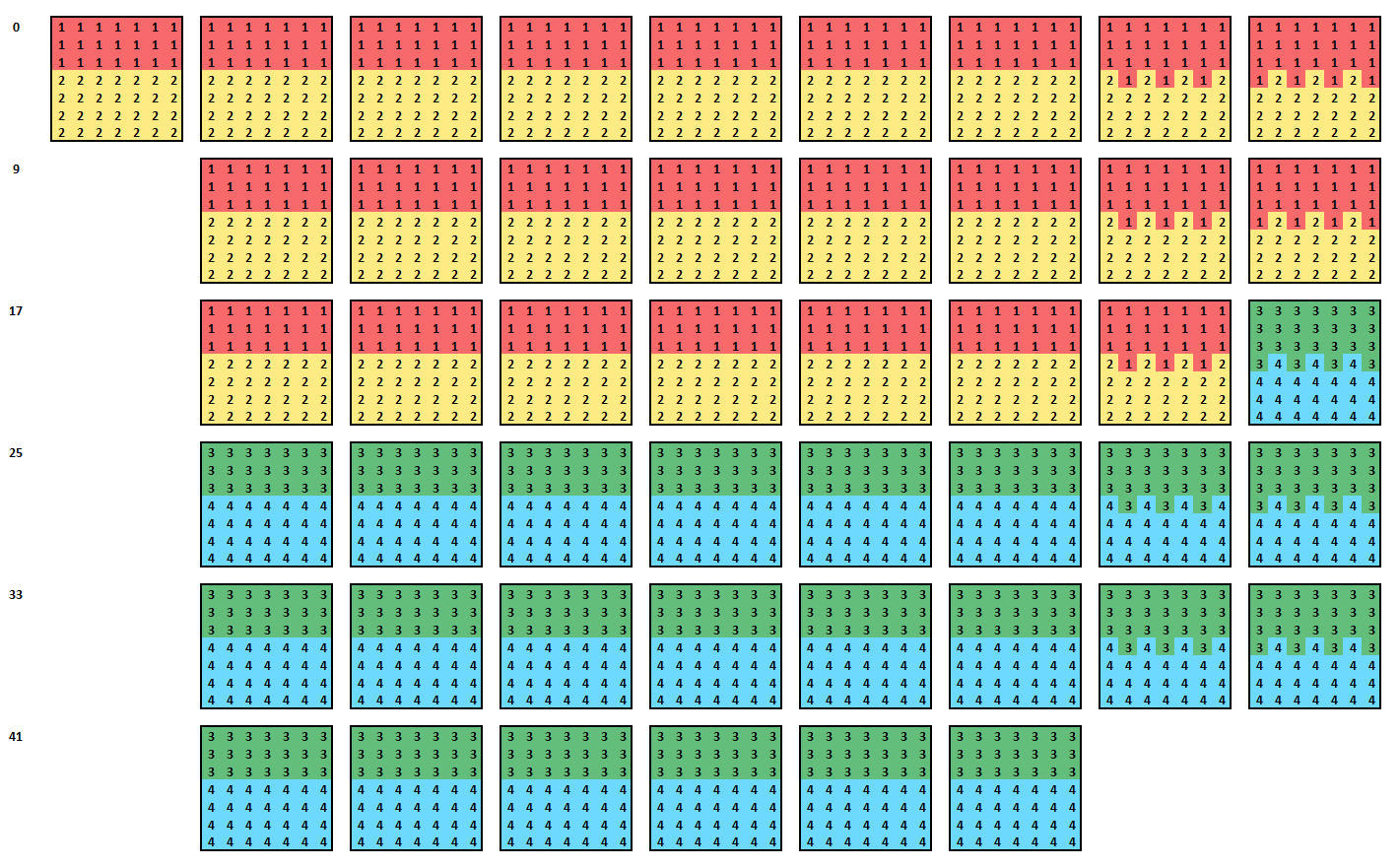}
  \caption{A 49-qubit, depth 46 circuit (left) and its
    partitioning (right).}
  \label{fig:googlesingleamp}
\end{figure}

We now describe a way to compute single amplitudes of universal random
circuits with $\sim 50$ qubits and depth $> 40$, which until very
recently was thought to be out of reach for current technology. We
discuss here the calculation of single amplitudes for a $7 \times
7$-qubit, depth 46 universal random circuit. As discussed in the
example above, we partition the circuit into four subcircuits: two
depth-23 subcircuits simulated in an input-to-output fashion, which
correspond to the first 23 layers and layer 0, and two depth-23
subcircuits simulated output-to-input, which correspond to the
remaining 23 layers. Both pairs of subcircuits are partitioned at the
boundary between the third and fourth rows of qubits, with the $CZ$
gates that bridge this boundary assigned to the top subcircuit in each
pair.  This partitioning is illustrated in
Fig.~\ref{fig:googlesingleamp}.

The four tensors constructed for these four subcircuits will
incorporate various subsets of indices.  One of these subsets of
indices represents the quantum state of the qubits after applying the
gates in layers 0 through 23.  Let $q_1,\ldots,q_{49}$ be these
indices with qubits numbered in left-to-right, top-to-bottom order.
Fourteen additional entanglement indices also need to be introduced in
the tensors representing layers 0 through 23 to account for the
deferred contraction of the $CZ$ gates that bridge the third and
fourth rows in layers 7, 8, 15, and 16.  Let $e_1,\ldots,e_{14}$ be
these indices.  Similarly, 14 additional entanglement indices need to
be introduced in the tensors representing the remaining 23 layers to
account for the deferred contraction of the $CZ$ gates that bridge the
third and fourth rows in layers 31, 32, 39, and 40.  Let
$e_{15},\ldots,e_{28}$ be these indices.  Because $CZ$ gates are
diagonal, additional entanglement indices are not needed for the
bridging $CZ$ gates at depths 23 and 24.  Instead,
$q_{22},\ldots,q_{28}$ serve as these entanglement indices.

Following the nomenclature used in the above example, the four tensors
calculated from the four subcircuits are then:
\begin{enumerate}
\item $\phi_{q_{1}\cdots q_{21}e_{1}\cdots e_{14}q_{23}q_{25}q_{27}}$
  corresponding to the ``top-left'' subcircuit, requiring 4~TB
  of storage in double precision;
\item $\xi_{q_{22}\cdots q_{49}e_{1}\cdots e_{14}}$ corresponding to
  the ``bottom-left'' subcircuit, requiring 64~TB of storage in
  double precision;
\item $\psi_{q_{1}\cdots q_{21}e_{15}\cdots
  e_{28}q_{22}q_{24}q_{26}q_{28}}$ corresponding to the ``top-right''
  subcircuit, requiring 8~TB of storage in double precision;
\item $\chi_{q_{22}\cdots q_{49}e_{15}\cdots e_{28}}$ corresponding to
  the ``bottom-right'' subcircuit, requiring 64~TB of storage in
  double precision;
\end{enumerate}
Note that the total memory requirement of 140~TB for these four
tensors lies well within the RAM limits of existing supercomputers.  The
horizontal-first contraction of the two top tensors and the two
bottom tensors yields two intermediate tensors of 0.5~TB each,
and the final amplitude calculation amounts to a complex-valued dot
product:
\[\begin{array}{rl}
  \sigma_{q_{22}\cdots q_{28}e_{1}\cdots e_{28}} &= \sum_{q_{1},\ldots,q_{21} \in \{0,1\}}
    \phi_{q_{1}\cdots q_{21}e_{1}\cdots e_{14}q_{23}q_{25}q_{27}}
    \psi_{q_{1}\cdots q_{21}e_{15}\cdots  e_{28}q_{22}q_{24}q_{26}q_{28}} \\
  \omega_{q_{22}\cdots q_{28}e_{1}\cdots e_{28}} &= \sum_{q_{29},\ldots,q_{49} \in \{0,1\}}
    \xi_{q_{22}\cdots q_{49}e_{1}\cdots e_{14}}
    \chi_{q_{22}\cdots q_{49}e_{15}\cdots e_{28}} \\
  \alpha &= \sum_{q_{22},\ldots,q_{28},e_{1},\ldots,e_{28} \in \{0,1\}}
    \sigma_{q_{22}\cdots q_{28}e_{1}\cdots e_{28}}
    \omega_{q_{22}\cdots q_{28}e_{1}\cdots e_{28}}.
\end{array}\]

Although we have not carried out the above simulations, expected
execution times can nevertheless be estimated in a safe way (i.e.,
overestimated) by first developing parallelization schemes for the
computations, then generating and testing benchmark codes that provide
upper bounds on the computational loads per processing node for each
subtask, and finally combining these execution times with estimated
communication times based on the communication patterns and data
transfer volumes dictated by the parallelization schemes. The
estimated run times based on this methodology are 17.93 hours on the
Vulcan supercomputer described earlier, and 4.76 hours on Sequoia, an
IBM Blue Gene/Q supercomputer also located at Lawrence Livermore
National Laboratory that is $4\times$ the size of Vulcan. To obtain
safe upper bounds on the expected runtime, the benchmark codes
responsible for 98.7\%\ of the estimated times did not employ
high-performance computing techniques such as cache blocking and loop
unrolling.  Faster run times should be achievable in actual,
well-engineered implementations.

The recent paper \cite{li2018quantum} describes the computation of a
single amplitude for a $7 \times 7$, depth 55 circuit, using the idea
of simulating the circuit in input-to-output order for approximately
half the total depth, and in output-to-input order for the remaining
layers. Once the two halves $L\ket{y}$ and $R^\dag \ket{x}$ of the
circuit are simulated, \cite{li2018quantum} computes the dot product
$\left(R^\dag \ket{x}\right)^{\dag} L \ket{y}$ in slices in the most
direct way, without changing the order of the tensor computation as
discussed in this paper to reduce resource consumption. Such
reordering does not appear to be necessary in the case of
\cite{li2018quantum}, likely because it uses a supercomputer (Sunway
TaihuLight, ranked 1st in the TOP500 list of supercomputers as of
November 2017) with significantly more memory than Vulcan or Sequoia.

\subsection{Leveraging secondary storage}
The paper \cite{haner2017simulation} suggests that solid-state
disk, or more generally secondary storage, could be used to supplement
main memory in order to simulate circuits whose quantum states are too
large to store in main memory alone. We combine the methods presented
here with those in \cite{haner2017simulation} to describe a viable
computation scheme that exploits secondary storage to simulate deeper
circuits than was thought possible.

The two methodologies are related by the fact that both involve
circuit partitioning and both employ tensor slicing.  In the case of
\cite{haner2017simulation}, ``global'' qubits used to index across
processing nodes correspond to tensor indices that are being sliced,
and ``local'' qubits correspond to tensor indices that are being used
to index into tensor slices stored on each processing node.  In
\cite{haner2017simulation}, circuits are partitioned so that all gates
within a subcircuit can be applied to update quantum state tensors on
a per-slice basis without communicating quantum state information
between processing nodes.  Zero-communication updates are possible
when all non-diagonal gates in a subcircuit are being applied to
``local'' qubits only.  Such updates are also possible for a handful
of additional circumstances described in \cite{haner2017simulation}.
In effect, circuits are partitioned by selecting different subsets of
``local'' qubits and analyzing which gates can be applied to yield
corresponding subcircuits.  During simulation, communication between
processing nodes occurs only when the simulation switches from one
subcircuit to another.  During these communication phases, the memory
layouts of quantum state tensors are reorganized so that new subsets
of indices are being used to ``globally'' index across processing
nodes versus ``locally'' index within the memories of individual
nodes, according to the needs of the next subcircuit to be simulated.

In the methods presented earlier, we considered circuit partitionings
in which the resulting tensors either fit in available aggregate
primary memory in their entirety, or slices of the resulting tensors
could be computed using available primary memory based on other
tensors already computed and stored in primary memory.  The resulting
tensors and/or their slices will generally be larger than the primary
memories of individual processing nodes, which represents a difference
in the way tensor slicing is being viewed in this paper as compared to
\cite{haner2017simulation}.  The techniques presented in
\cite{haner2017simulation} can be combined with those presented here
to employ secondary storage when quantum states are too large to fit
in aggregate primary memory.  Because secondary storage is typically
orders of magnitude slower than main memory, the viability of using it
depends on the extent to which the number of read/write cycles can be
minimized or overlapped with computation.  To achieve such
minimization, we first employ our decomposition ideas to partition the
initial portions of a circuit so as to maximize the number of gates
that can be simulated using available aggregate memory, with the
resulting quantum state then calculated in slices and written to
secondary storage.  The partitioning methods discussed in
\cite{haner2017simulation} can then be applied to the remaining gates
in the circuit with the number of ``local'' qubits set higher,
according to the size of aggregate memory instead of the memory size
available on individual processing nodes.  The resulting tensor slices
will then be much larger, allowing many more gates to be simulated
before additional secondary-storage read/write cycles are needed. The
resulting subcircuits can then be further partitioned into
sub-subcircuits to minimize internode communication in the overall
calculations.

We now provide a specific implementation of the above idea and
estimate its computational cost. Simulating universal random circuits
of $\sim 50$ qubits with arbitrary depth using secondary storage
results in higher execution times due to the relatively high cost of
disk read/write operations. However, we find both that the slowdown is
considerably less than twofold for the instances under study, and that
recent system advancements, such as NVRAM-based burst buffers, can
have a highly beneficial effect on these run times.  The larger memory
pool available via secondary storage allows us to push the boundary of
quantum circuit simulations even further. In order to do so, we
observe that, by construction, universal random circuits have groups
of qubits on the boundary of the grid that periodically do not
interact with other qubits for several layers of gates. In particular,
for circuits on a $7 \times 7$ grid, a group of $7$ qubits at the
boundary has two layers of two-qubit interactions with other rows or
columns, followed by six layers without further interactions (only
single-qubit gates or two-qubit gates within the group). For this
reason, slicing these qubits is very effective: we can choose one of
the boundary rows or columns, slice the corresponding qubits for six
layers, and simulate the circuit applied to the remaining qubits
independently.  Recently, \cite{haner2017simulation} used a similar
observation to decide how to distribute computation across processors.

\begin{figure}[tb]
  \centering
  \includegraphics[width=0.7\textwidth]{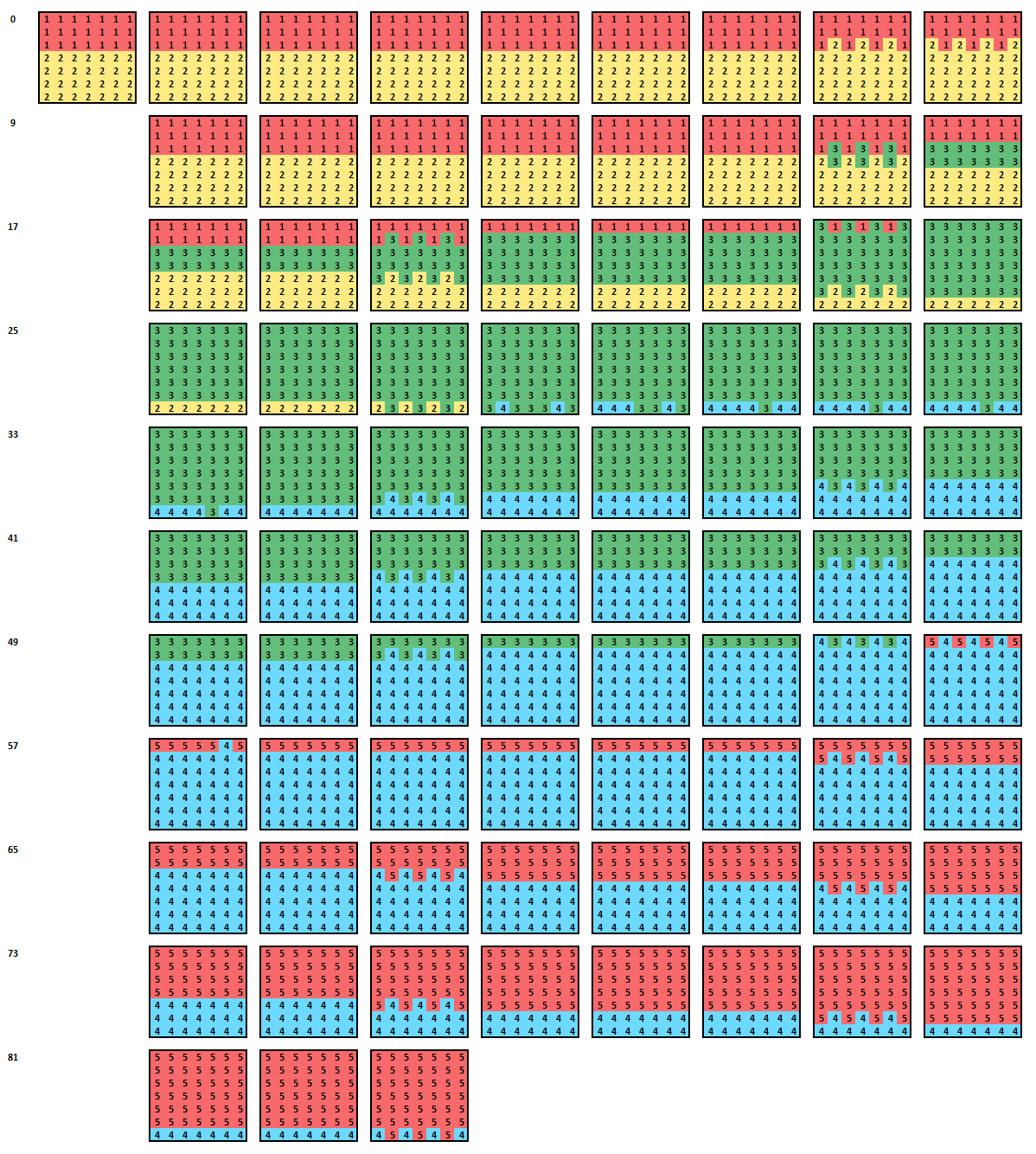}
  \caption{Partitioning of a 49-qubit, depth 83 circuit.}
  \label{fig:googlewithstorage}
\end{figure}

Our strategy consists in the following. We slice qubits at the
boundary of the grid (e.g., the bottom row of the grid), simulating
the remaining part of the circuit using the Schr\"odinger method with a
tensor as large as our memory allows. The simulation is carried
out for as many layers of gates as possible without introducing
additional entanglement indices. Remarkably, for qubits on the
opposite row/column (e.g., the top row of the grid) this allows us to
apply over 30 layers of gates, whereas for qubits closer to the sliced
qubits we can only progress for a handful of layers before we are
forced to stop or increase memory occupancy. This produces several
slices of the same size, which are stored to disk. To apply further
gates we start a new subcircuit, slicing qubits in the row/column
opposite to those sliced in the previous subcircuit (e.g., the top row
of the grid), and simulating the rest of the circuit with the
Schr\"odinger method. The initial state can be loaded from the
slices stored on disk, and the process can be iterated. This yields a
``wave'' pattern in the subcircuits: each subcircuit starts from the
top or bottom row of qubits, then extends to the rest of the circuit
until it encompasses the row opposite to the starting set of qubits,
at which point it shrinks back. We show the proposed circuit
partitioning in Fig.~\ref{fig:googlewithstorage}. The specific example
of Fig.~\ref{fig:googlewithstorage} uses a circuit with depth 83, but
the pattern used for subcircuits 4, 5 and 6 can be carried out
indefinitely, adding further disk read/write operations. In the
language of Alg.~\ref{alg:sim}, there are 5 subcircuits:
$V_1,\dots,V_5$ labeled $1,\dots,5$ in
Fig.~\ref{fig:googlewithstorage}; we assign $\sigma_1 = \sigma_2 = 1,
\sigma_3 = 2, \sigma_4 = 3, \sigma_5 = 4$, with precedence $V_1 \prec
V_3, V_2 \prec V_3, V_3 \prec V_4, V_4 \prec V_5$. For every
subcircuit $V_i, i \ge 3$, a row of qubits is sliced because the
corresponding hyperedges connect $V_{i-1}$ to $V_{i+1}$ (if any
non-diagonal single-qubit gates are applied on these qubits, instead
of just one hyperedge per qubit we could have several shorter ones,
which we would also slice).

For circuits of this size, it is important that the simulation
algorithm is well-engineered in all of its aspects. Thus, we describe
here the implementation choices that we used to estimate run times. We
discuss a depth-83 circuit; the tables provide data for a depth-55
circuit as well.

Tensors for $V_1$ and $V_2$ in Fig.~\ref{fig:googlewithstorage} are
the first to be calculated, and the resulting pair of tensors is
contracted one slice at a time. The slicing process proceeds by
looping over the possible values for qubits 43--49 and slicing the
tensor for $V_2$ on these values prior to performing the contraction
with the tensor for $V_1$.  For each of the 128 resulting slices, the
gates belonging to $V_3$ in Fig.~\ref{fig:googlewithstorage} are
applied to the contraction results and the resulting updated slices
are transferred to secondary storage.  This process of slicing,
contracting, applying gates, and sending results to secondary storage
is repeated 128 times, once for each of the 128 possible values of
qubits 43--49. The gates belonging to $V_4$ in
Fig.~\ref{fig:googlewithstorage} are applied to the intermediate
results that were transferred to secondary storage.  These gate
applications can also be performed in slices, this time slicing on
qubits 1--7.  To ensure the retrieval from secondary storage is
performed efficiently, data can be organized in secondary storage as
$2^{14}$ logical files indexed by the values of qubits 1--7 and
43--49, wherein each logical file contains $2^{35}$ complex amplitudes
corresponding to qubits 8--42.  Thus, in the phase discussed above of
applying gates in $V_3$, for each of the 128 values of qubits 43--49
that are being sliced, 128 logical files are written to secondary
storage, corresponding to the 128 possible values of qubits 1--7.  In
the phase that applies the gates in $V_4$, for each of the 128 values
of qubits 1--7 that are being sliced, 128 logical files are read from
secondary storage, corresponding to the 128 possible values of qubits
43--49.  Once these 128 files of amplitudes are loaded into memory,
the gates in $V_4$ can be applied and each updated slice can be
written back to storage; alternatively, the final amplitudes can be
processed in-memory on a per-slice basis.

For a depth-55 circuit, the simulation process ends with $V_4$ (in
fact, we only need to apply the gates in $V_4$ at depth $\le 55$).  To
continue the simulation to depth 83, we repeat the process for $V_5$,
shown in Fig.~\ref{fig:googlewithstorage}, to simulate the remaining
gates.  We proceed in the same manner as for $V_4$, except this time,
for each of the 128 values of qubits 43--49 that are now being sliced,
128 logical files are loaded from secondary storage corresponding to
the 128 possible values of qubits 1--7.

The timing results in \cite{haner2017simulation} can be used to
estimate computation and communication times for the above
calculations, assuming that the parallelization methods discussed in
\cite{haner2017simulation} are being used and one is running on the
equivalent of 4096 nodes of a Cori-II-class supercomputer. These times
are reported in Tables \ref{tab:depth55time} (for depth 55) and
\ref{tab:depth83time} (for depth 83).

\begin{table}
  \centering
  \begin{tabular}{|c|c|c|c|c|}
    \hline
    & & Compute & All-to-All & \\
    & Num & Time & Time & Total \\
    Subcircuit & Gates & (hours) & (hours) & (hours) \\
    \hline
    1 & 215 & 0.00 & 0.00 & 0.00 \\
    2 & 336 & 0.00 & 0.01 & 0.01 \\
    Contract & N/A & 0.39 & 0.00 & 0.39 \\
    3 & 621 & 0.94 & 3.05 & 3.98 \\
    4 & 346 & 0.52 & 2.03 & 2.55 \\
    \hline
    Total & 1518 & 1.85 & 5.08 & 6.93 \\
    \hline
  \end{tabular}
  \caption{Estimated computation and communication times for a $7
    \times 7$-qubit, depth 55 random circuit simulated on the
    equivalent of 4096 nodes of a Cori-II-class supercomputer.  Time
    estimates do not include secondary storage access.}
  \label{tab:depth55time}
\end{table}
\begin{table}
  \centering
  \begin{tabular}{|c|c|c|c|c|}
    \hline
    & & Compute & All-to-All & \\
    & Num & Time & Time & Total \\
    Subcircuit & Gates & (hours) & (hours) & (hours) \\
    \hline
    1 & 215 & 0.00 & 0.00 & 0.00 \\
    2 & 336 & 0.00 & 0.01 & 0.01 \\
    Contract & N/A & 0.39 & 0.00 & 0.39 \\
    3 & 618 & 0.93 & 3.05 & 3.98 \\
    4 & 756 & 1.14 & 3.05 & 4.19 \\
    5 & 348 & 0.53 & 2.03 & 2.56 \\
    \hline
    Total & 2273 & 2.99 & 8.13 & 11.12 \\
    \hline
  \end{tabular}
  \caption{Estimated computation and communication times for a $7
    \times 7$-qubit, depth 83 random circuit simulated on the
    equivalent of 4096 nodes of a Cori-II-class supercomputer.  Time
    estimates do not include secondary storage access.}
  \label{tab:depth83time}
\end{table}

The estimated compute times for $V_1$ and $V_2$ presented in
Tables~\ref{tab:depth55time} and~\ref{tab:depth83time} are obtained
by scaling the $6 \times 5$-qubit simulation times reported in Table~1
in \cite{haner2017simulation} according to the number of gates.  The
tensor values for $V_1$ and $V_2$ can be calculated in an
embarrassingly parallel fashion without communication, where the local
tensors in each node have rank 28.  Thus, a single-node, 30-qubit
simulation provides an upper bound to these compute times.

Estimated compute times for all other rows in
Tables~\ref{tab:depth55time} and~\ref{tab:depth83time} are obtained by
using the percent-communication column reported in Table~1 in
\cite{haner2017simulation} to first decompose the reported simulation
time for a $7 \times 6$-qubit circuit into a per-gate compute time and
a per-all-to-all communication and synchronization time.  This
per-gate compute time is scaled by multiplying by the number of gates in
each of the subcircuits under consideration, and then further
multiplying by the number of slices (i.e., 128).  The $7 \times 6$
timing results reported in \cite{haner2017simulation} are used because
subcircuits 3, 4, and 5 in Fig.~\ref{fig:googlewithstorage} correspond
to $7 \times 6$-qubit subcircuits and the tensor slices involved in
their simulation have rank 42. The sliced tensor contraction results
for $V_1$ and $V_2$ also have rank 42.  These tensor contractions
require 128 complex multiplies and 127 complex additions per
amplitude.  An upper bound to the compute time is estimated by
assuming an equivalent circuit-size of 256 gates to model the
execution time.

All-to-all communication and synchronization times are estimated using
the decomposed per-all-to-all time discussed above together with
Figure~5 in \cite{haner2017simulation} to determine the number of
all-to-all communication cycles that are needed as a function of the
depths of each subcircuit.  Accordingly, for a depth 55 circuit, $V_3$
should require three all-to-alls per slice, while $V_4$ should require
only two all-to-alls per slice.  Similarly, for a depth 83 circuit,
$V_3$ and $V_4$ should each require three all-to-alls per slice, while
$V_5$ should require only two all-to-alls per slice.  In addition, one
all-to-all would be needed after computing the tensor for $V_2$ to
redistribute it in preparation for the contraction of $V_1$ and $V_2$.
Each of these sources of all-to-all communication are reflected in
Tables~\ref{tab:depth55time} and~\ref{tab:depth83time}.

\begin{table}
  \centering
  \begin{tabular}{|l|c|c|c|c|}
    \hline
    & Size & Transfer Rate & Single Precision & Double Precision \\
    Storage System & (PB) & (TB/sec) & (hours) & (hours) \\
    \hline
    Summit Burst Buffer & 7 & 10.00 & 0.23 & \\
    Summit File System & 250 & 2.20 & 1.03 & 2.07 \\
    Sequoia File System & 50 & 0.83 & 2.74 & 5.48 \\
    Vulcan File System & 5 & 0.11 & 21.18 & \\
    \hline
  \end{tabular}
  \caption{Secondary storage sizes available on certain IBM
    supercomputers, the corresponding transfer rates, and the times
    needed to write $2^{49}$ quantum amplitudes to secondary storage
    and then read them back assuming full sustained transfer
    rates. Summit is the IBM-Power9/NVIDIA-Volta supercomputer
    at the Oak Ridge National Laboratory.}
  \label{tab:readwritetime}
\end{table}

\begin{table}
  \centering
  \begin{tabular}{|l|c|c|c|c|c|c|}
    \hline
    & & & Single & Double & Single & Double \\
    & & & Precision & Precision & Precision & Precision \\
    & Compute & All-to-All & Read/Write & Read/Write & Total & Total \\
    Storage System & (hours) & (hours) & (hours) & (hours) & (hours) & (hours) \\
    \hline
    Summit Burst Buffer & 1.85 & 5.08 & 0.23 & -- & 7.16 & -- \\
    Summit File System & 1.85 & 5.08 & 1.03 & 2.07 & 7.97 & 9.00 \\
    Sequoia File System & 1.85 & 5.08 & 2.74 & 5.48 & 9.68 & 12.42 \\
    Vulcan File System & 7.40 & 20.33 & 21.18 & -- & 48.92 & --  \\
    \hline
  \end{tabular}
  \caption{Estimates of total run times, including secondary storage
    read/write times, for the computation of the full state vector of
    a $7 \times 7$, depth-55 circuit.}
  \label{tab:depth55}
\end{table}

\begin{table}
  \centering
  \begin{tabular}{|l|c|c|c|c|c|c|}
    \hline
    & & & Single & Double & Single & Double \\
    & & & Precision & Precision & Precision & Precision \\
    & Compute & All-to-All & Read/Write & Read/Write & Total & Total \\
    Storage System & (hours) & (hours) & (hours) & (hours) & (hours) & (hours) \\
    \hline
    Summit Burst Buffer & 2.99 & 8.13 & 0.46 & -- & 11.58 & -- \\
    Summit File System & 2.99 & 8.13 & 2.07 & 4.14 & 13.19 & 15.26 \\
    Sequoia File System & 2.99 & 8.13 & 5.48 & 10.97 & 16.60 & 22.09 \\
    Vulcan File System & 11.97 & 32.52 & 42.37 & -- & 86.85 & -- \\
    \hline
  \end{tabular}
  \caption{Estimates of total run times, including secondary storage
    read/write times, for the computation of the full state vector of
    a $7 \times 7$, depth-83 circuit.}
  \label{tab:depth83}
\end{table}

Tables~\ref{tab:depth55}-\ref{tab:depth83} shows estimated overall run
times, obtained by combining Tables \ref{tab:readwritetime},
\ref{tab:depth55time} and \ref{tab:depth83time}. We remark that these
estimates assume the equivalent of 4096 of Cori II's 9304 compute
nodes. Because Sequoia and Summit are both ranked higher than Cori II
in terms of their High Performance Linpack benchmarks, these execution
times should be achievable on both Sequoia and Summit, and they can
likely be improved upon.  Vulcan is four times smaller than Sequoia, and
so the computation and communication times estimated for Sequoia were
multiplied by four to produce entries for Vulcan.  The above estimates
are thus very safe.  The secondary storage transfer time estimates are
less safe, being solely based on published data transfer rates.

It should be noted that the overall run time estimates in
Tables~\ref{tab:depth55}-\ref{tab:depth83} are all dominated by the
computation and communication times, and not the secondary storage
transfer times.  Because these secondary storage transfer times are
not dominant, we can conclude that secondary storage provides a viable
basis for simulating quantum circuits to arbitrary depth when quantum
states are too large to fit in main memory alone. This concludes our
analysis of computing times for deep circuits.

The preceding analysis provides several insights that can be
  used to arrive at an automated process for partitioning circuits to
  leverage secondary storage. The first is that data organization on
  secondary storage should remain fixed during simulation, so that
  previously stored amplitudes can simply be overwritten with updated
  amplitudes without incurring additional overhead.  A second insight
  is that, in order to make such overwriting as efficient as possible,
  secondary storage should be organized into tensor slices (i.e., the
  logical files discussed above) in such a way that the ``global''
  qubits used to index into secondary storage are always a superset of
  the ``global'' qubits used to slice the computations.  In the above
  example, qubits 1--7 and 43--49 were used to index into secondary
  storage while the computations alternated between slicing on qubits
  43--49 and on qubits 1--7.

We now describe a greedy optimization approach that can be
  employed to identify such supersets of ``global'' qubits while at
  the same time trying to maximize the number of gates processed in
  each read/write cycle.  The latter is essentially a proxy to
  minimizing the number of read/write cycles.  The optimization
  proceeds in two stages.

The first stage consists of applying a suitable in-memory
  circuit partitioning method (e.g., the $A^{*}$ or integer
  programming approaches presented in this paper) to depth-truncated
  versions of the circuit to be simulated, in order to determine the
  maximum depth at which the circuit can be simulated entirely
  in-memory.  The qubits that are identified for slicing at this
  maximum depth are then used as the ``global'' qubits for the first
  phase of simulation prior to writing to secondary storage.  In
  determining which qubits can be sliced, single-qubit diagonal gates
  are treated as non-diagonal for the purpose of identifying slicing
  opportunities.  The rationale is that subsequent stages of analysis,
  described below, do not make use of these properties, so, in order to
  maintain consistency, we neglect to leverage these properties in the
  first stage as well. The properties of single-qubit diagonal gates can be
  exploited to optimize the computations entailed by the first stage
  of circuit partitioning, but we do not consider them in the choice
  of ``global'' qubits that will determine the organization of
  secondary storage.  The final step of this first stage of
  optimization is to analyze the circuit partitionings obtained by the
  above approach to determine which of the remaining gates in the
  circuit can be applied without requiring any additional memory or
  contractions between subcircuits.

For example, suppose that the circuits shown in
  Fig.~\ref{fig:googlepart} actually extend much deeper and that a
  circuit partitioning method determined that the $7 \times 7$ and $8
  \times 7$ circuits could only be simulated in-memory to depths 27
  and 23, respectively, using the illustrated partitionings.  In each
  case, according to the discussion above, the qubits interior to
  subcircuit 3 would not be sliced in the manner illustrated in
  Fig.~\ref{fig:googlepart} because the slicing of these qubits relies
  on the presence of corresponding single-qubit diagonal gates in
  subcircuit 3 of each circuit.  Instead, for the $7 \times 7$ circuit,
  only qubits 43--49 would be sliced, and for the $8 \times 7$ circuit,
  qubits 1--7 and 50--56 would be sliced.  When these subcircuits are
  then extended by adding as many subsequent gates as possible,
  subcircuit 3 of the $7 \times 7$ circuit would be extended as deep
  as depth 55, as illustrated in Fig.~\ref{fig:googlewithstorage}.
  Similarly, subcircuit 3 of the $8 \times 7$ circuit would be
  extended as deep as depth 39 (not illustrated), but the resulting
  extension of subcircuit 3 would form a diamond-shaped pattern
  instead of the triangular patterns that can be seen in
  Fig.~\ref{fig:googlewithstorage} for the $7\times 7$ circuit.  Note
  that in both cases, slicing qubits interior to subcircuit 3 to
  exploit single-qubit diagonal gates would have interfered with these
  subcircuit extensions.

The second stage of optimization selects ``global'' qubits for
  subsequent circuit partitions in a greedy fashion that attempts to
  maximally extend the depth of simulation at each phase.  The
  analysis proceeds by first identifying, for each qubit, the depth of
  the deepest gate applied to that qubit given the circuit
  partitioning constructed thus far.  The greedy heuristic is to slice
  on those qubits that correspond to the deepest of these gates,
  choosing a sufficient number of such ``global'' qubits to enable the
  per-slice computations for the resulting subcircuit to be performed
  in available aggregate main memory.  Ties among possible choices are
  broken either randomly or by performing a greedy look-ahead search
  to evaluate which subset of candidate ``global'' qubits leads to the
  greatest circuit depth and/or number of gates in the resulting
  subcircuit.

The heuristic of choosing ``global'' qubits according to the
  deepest gates guarantees that the corresponding ``local'' qubits for
  the resulting subcircuit will incorporate qubits associated with the
  shallowest gates in the previous subcircuit.  Empirically, these
  latter qubits tend to be associated with the deepest gates in the
  new subcircuit, which gives rise to the ``wave'' patterns observed
  in Fig.~\ref{fig:googlewithstorage} as discussed above.  For
  example, in Fig.~\ref{fig:googlewithstorage} qubits 43--49 are
  sliced in the first phase and they end up with the shallowest gates
  at depths 26 and 27, while qubits 1--7 have the deepest gates at
  depths 54 and 55.  Subsequently selecting and slicing on qubits 1--7
  in the second phase results in qubits 43--49 now having the deepest
  gates while qubits 1--7 have the shallowest gates.  The pattern then
  repeats.

The above process of choosing ``global'' qubits according to the
  deepest gates in successive partitioning phases is repeated until
  the circuit is completely segmented into subcircuits.  The union of
  the ``global'' qubits of these subcircuits becomes the set of
  ``global'' qubits used to organize secondary storage.  The ``local''
  qubits for secondary storage will thus always be a subset of the
  ``local'' qubits at each phase of computation, which implies that
  each tensor slice during a phase of computation corresponds to a set
  of tensor slices on secondary storage.  This correspondence further
  implies that during simulation, appropriate sets of tensor slices on
  secondary storage are loaded into aggregate main memory,
  operated upon, and their updated values are then written back to
  secondary storage without affecting any of the other tensor slices
  in secondary storage.  The proposed approach minimizes secondary
  storage traffic during each computational phase and the total number
  of computational phases, thus minimizing total secondary storage access,
  through good choices of ``global'' qubits and circuit
  partitionings.

Note that it is conceivable that there exist circuit topologies for
  which the above process might yield a union of ``global'' qubits
  that incorporates all or too many qubits, thereby making slice sizes
  too small to enable efficient secondary storage transfer.  In this
  case the selection process must be constrained.  The simplest
  (greedy) approach would be to set a threshold on the maximum number
  of ``global'' qubits (or equivalently on the minimum number of
  ``local'' qubits) that can be used to organize secondary storage and
  to then choose ``global'' qubits for circuit partitioning in the
  same manner as before, but with the choice of deepest gates
  restricted to previously selected ``global'' qubits once this
  threshold is met.

\subsection{On the graph representation of two-qubit gates}
The tensor network representation used in our paper differs from
\cite{markov2008simulating} because of the presence of hyperedges. As
discussed in the main text, this leads to a more accurate computation
of the size of the tensors for diagonal gates, and allows more
efficient circuit graph decompositions. As it turns out, our
representation is equivalent to that of
\cite{boixo2017simulation}. This is now explained in more detail.

We consider a two-qubit diagonal gate, since this is the only case in
which differences between our paper and \cite{markov2008simulating}
arise. This difference is also the reason we sometimes choose to
transform $CX$ gates into $CZ$ gates and Hadamards, to take advantage
of the more efficient treatment of the diagonal $CZ$ gates. The
complexity of the simulation algorithm described in
\cite{markov2008simulating} depends on the treewidth of the line graph
of the circuit graph. Given a hypergraph $G = (V,E)$, its line graph
$G^* = (V^*, E^*)$ is defined by the vertex set $V^* := \{e \in E\}$
and the edge set $E^* := \{(e^1, e^2) \in V^* \times V^* : e^1 \cap
e^2 \neq \emptyset\}$. Notice that this definition matches the
traditional definition of line graphs when applied to a regular graph
$G$ (as opposed to a hypergraph). We show the difference between the
line graph of our representation of the graph circuit, and that of
\cite{markov2008simulating}, using the circuit given in
Fig.~\ref{fig:linegraphcircuit}. The hypergraph of the circuit in
Fig.~\ref{fig:linegraphcircuit} has only four hyperedges because there
are four different index labels in the circuit, while the traditional
tensor network representation has eight --- one for each wire
segment. This leads to a 4-clique in the line graph of the traditional
tensor network (Fig.~\ref{fig:linegraph}, left), while the line graph
of the hypergraph has only 2-cliques (Fig.~\ref{fig:linegraph},
right). The reader can easily verify that the graphical model proposed
in \cite{boixo2017simulation} in the context of a variable elimination
algorithm yields exactly the same graph representation of the line
graph of the hypergraph. This is due to the fact that both our
hypergraph and the variable elimination model put more emphasis on the
indices of the tensor network than in \cite{markov2008simulating}.

\begin{figure}[tb]
  \leavevmode
  \centering
  \Qcircuit @C=1em @R=.7em {
    \lstick{\ket{0}} & \ustick{i_0} \qw & \gate{T} & \ustick{i_0} \qw & \ctrl{1}  & \ustick{i_0} \qw & \gate{T} & \ustick{i_0} \qw   \\
    \lstick{\ket{0}} & \ustick{i_1} \qw & \gate{H} & \ustick{j_1} \qw & \gate{Z}  & \ustick{j_1} \qw & \gate{H} & \ustick{k_1} \qw 
  }
  \caption{Example circuit for the line graph discussion.}
  \label{fig:linegraphcircuit}
\end{figure}

\begin{figure}[tb]
  \centering
  \includegraphics[width=0.7\textwidth]{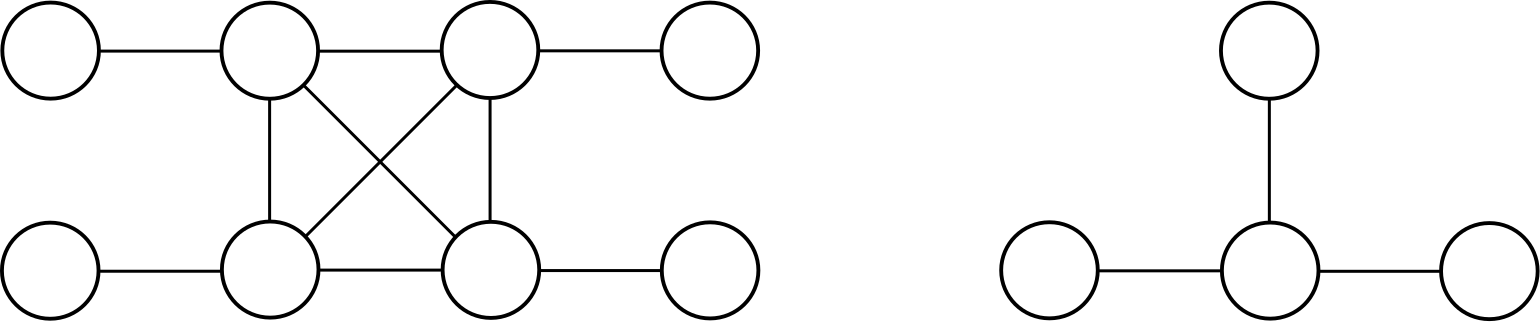}
  \caption{Line graph of the circuit in
    Fig.~\ref{fig:linegraphcircuit}, using traditional tensor networks
    (left), and using the graphical model advocated in this paper
    (right).}
  \label{fig:linegraph}
\end{figure}



\bibliography{breaking_49qubit}

\bibliographystyle{abbrv}

\end{document}